\documentclass[prd, twocolumn, showpacs, showkeys, nofootinbib,superscriptaddress]{revtex4-1}

\usepackage{amssymb,amsmath,bbm}
\usepackage{amsmath}
\usepackage{graphicx}
\usepackage[usenames,dvipsnames]{color}                                
\usepackage{hyperref}
\usepackage{comment}                                                   
\usepackage{subfig}
\usepackage{enumerate}
\hypersetup{                 
    colorlinks=true,         
    linkcolor=blue,          
    citecolor=blue,           
    urlcolor=Violet          
}

\let\oldmarginpar\marginpar
\renewcommand\marginpar[1]{\oldmarginpar{\color{red}\raggedright\scriptsize #1}}

\newcommand{\mean}[1]{\ensuremath{\lf\langle #1 \rt\rangle }}
\newcommand{\abs}[1]{\ensuremath{\lf| #1 \rt| }}
\newcommand{\diby}[2]{\ensuremath{\frac{\partial #1}{\partial #2}}}

\def\lf {\ensuremath{\left}}
\def\rt {\ensuremath{\right}}

\def\de {\ensuremath{ {\rm d} }}

\begin{document}

\title{\textbf{\Large Bouncing Unitary Cosmology\\I. Mini-Superspace  General Solution}}

\author{Sean Gryb}\email{sean.gryb@gmail.com}
\affiliation{{{\it Department of Philosophy}, University of Bristol}}
\affiliation{{{\it H. H. Wills Physics Laboratory}, University of Bristol}}
\author{Karim P. Y. Th\'ebault}\email{karim.thebault@bristol.ac.uk}
\affiliation{{{\it Department of Philosophy}, University of Bristol}}

\date{\today}


\begin{abstract}
    We offer a new proposal for cosmic singularity resolution based upon a quantum cosmology with a unitary bounce. This proposal is illustrated via a novel quantization of a mini-superspace model in which there can be superpositions of the cosmological constant. This possibility leads to a finite, bouncing unitary cosmology. Whereas the usual Wheeler--DeWitt cosmology generically displays pathological behaviour in terms of non-finite expectation values and non-unitary dynamics, the finiteness and unitarity of our model are formally guaranteed. For classically singular models with a massless scalar field and cosmological constant, we show that well-behaved quantum observables can be constructed and generic solutions to the universal Schr\"{o}dinger equation are singularity-free. Generic solutions of our model displays novel features including: i) superpositions of values of the cosmological constant; ii) universal effective physics due to non-trivial self-adjoint extensions of the Hamiltonian; and iii) bound `Efimov universe' states for negative cosmological constant. The last feature provides a new platform for quantum simulation of the early universe. A companion paper provides detailed interpretation and analysis of particular cosmological solutions that display a cosmic bounce due to quantum gravitational effects, a well-defined FLRW limit far from the bounce, and a semi-classical turnaround point in the dynamics of the scalar field which resembles an effective inflationary epoch. 

\end{abstract}

\maketitle
\tableofcontents



		


\section{Introduction}
\label{intro}

The `big-bang' cosmic singularity can be characterised classically in terms of both the incompleteness of causal (i.e., non-spacelike) past-directed curves and the existence of a curvature pathology \cite{penrose:1965,hawking:1970,hawking:1973,senovilla:1998}. That the pathology in question is physically problematic is unambiguously demonstrated by the existence of scalar curvature invariants that grow without bound in finite proper time along the incomplete curves in question \cite{thorpe:1977,curiel:1999}. Such pathological behaviour in observable classical quantities can be understood to signal the breakdown of general relativity and, thus, the requirement for new theoretical tools. A reasonable hope is that the cosmic singularity problem might be resolved via the introduction of inflationary mechanisms \cite{guth1981,linde:1982}. In particular, one might expect that eternal inflation models could resolve the initial singularity \cite{vilenkin:1983}. However, the Penrose-Hawking singularity theorems can, in fact, be extended to show that a broad range of `physically reasonable' eternal inflationary spacetimes are necessarily geodesically past incomplete and therefore singular in the relevant sense  \cite{borde:1994}. There are thus reasons to expect a `big-bang' singularity to exist \textit{prior} to the inflationary phase of such models. A complete description of the early universe requires us to introduce new physics that goes beyond inflation.   

Most contemporary approaches to resolving the cosmic singularity problem are based upon cosmic bounce scenarios. In such models the classical singularity is replaced by a new pre-big bang $t\rightarrow-\infty$ epoch. Big bounce cosmologies have hitherto been proposed based upon stringy effects \cite{brandenberger:1989,gasperini:1993,khoury:2001,finelli:2002,nayeri:2006}, path integral techniques \cite{hartle:1983,gielen:2016}, loop approaches \cite{bojowald:2001,ashtekar:2006,ashtekar:2006b,ashtekar2007loop}, and group field theory \cite{gielen:2016b,deCesare:2017,oriti:2017}. Whilst, bouncing cosmologies can be combined with inflation -- giving mixed scenarios -- a particular attraction is that the big bounce can potentially replace  inflation as the mechanism for solving many of the problems of standard big bang cosmology \cite{brandenberger:2017}. The horizon problem, in particular, is automatically solved in bouncing cosmologies, since the observed isotropy of the cosmic microwave background can be explained simply by causal interactions in the pre-big bang epoch. Bouncing cosmologies are thus attractive as either a supplement to or replacement of the inflationary cosmological paradigm. 

In this paper we propose an alternative quantization procedure for cosmology in which superpositions of the cosmological constant are allowed. This possibility allows for a new model of bouncing cosmology: a bouncing unitary cosmology. Our approach is based upon canonical quantization of isotropic and homogeneous mini-superspace models but differers significantly from the two standard canonical treatments of such models. The first, older approach is based upon a Dirac quantization of mini-superspace expressed in terms of ADM variables \cite{deWitt:equation,Misner:1969}, and leads to a Wheeler--DeWitt mini-superspace quantum cosmology. Simple models, where one considers a massless scalar field with zero spatial curvature and non-negative cosmological constant, can be shown explicitly to contain observable operators with divergent expectation values \cite{ashtekar:2008}. In this strong sense, the big bang singularity is not resolved in even the most basic class of mini-superspace quantum cosmologies.\footnote{Although there are good general reasons to expect singularities to persist in Wheeler--DeWitt mini-superspace cosmologies, forms of singularity resolution have been shown to hold for models with Brown-Kucha\v{r} dust fields \cite{amemiya:2009,lawrie:2012}. See \cite{bojowald:2007,Kiefer:2007,falciano:2015} and \S\ref{sec:sing res} for further discussion.} The second, newer approach is inspired by the techniques used in Loop Quantum Gravity and is known as Loop Quantum Cosmology \cite{bojowald:2001,ashtekar:2006,ashtekar:2006b,ashtekar2007loop}. Loop Quantum Cosmology relies upon a `polymer' representation of the observable algebra of mini-superspace  and involves the introduction of a Planck-scale cutoff for the problematic operators of the Wheeler--DeWitt theory. The same models that exhibit pathological behaviour under the Wheeler--DeWitt treatment can be shown to feature operators with finite expectation values when treated in Loop Quantum Cosmology \cite{ashtekar:2008}. In this precise sense, the big bang singularity is resolved in Loop Quantum Cosmologies for models where the Wheeler--DeWitt treatment fails.

In this paper, we will offer a new proposal for singularity resolution in mini-superspace quantum cosmology that is based upon evolution. Our method does not make use of polymer methods. Rather, the observable operators in our theory evolve unitarily and remain finite because they are `protected' by the uncertainty principle. The unitary evolution of operators in our model is based upon a novel approach to the quantization of ADM mini-superspace that leads to a Schr\"{o}dinger-type equation for the universe \cite{Gryb:2016a}. An equation of the same form was in fact derived some time ago in the context of unimodular gravity \cite{unruh:1989}. That earlier treatment did not include a detailed analysis of generic or specific cosmological solutions, an explicit construction of the observable operators, or an investigation of the fate of the singularity. More problematically, the quantization did not include the self-adjoint representation of the Hamiltonian necessary to guarantee unitarity via appeal to Stone's theorem. In our approach, the Hamiltonian will be given an explicit self-adjoint representation. Furthermore, for classically singular models with a massless scalar field and cosmological constant $\Lambda$, well-behaved quantum observables will be constructed and generic solutions to our `universal Schr\"{o}dinger equation' will be shown to be singularity-free. Key features of the solutions of our model include a cosmic bounce due to quantum gravitational effects, a well-defined FLRW limit far from the bounce, and a semi-classical turnaround point in the quantum dynamics of the scalar field which resembles an effective inflationary epoch. Our bounce scenario is phenomenologically distinct form those studied in the literature -- e.g. those found in LQC \cite{bojowald:2001}. In particular, our model displays novel features including: i) superpositions of values of the cosmological constant; ii) a non-zero scattering length around the big bounce; and iii) bound `Efimov universe' states for negative cosmological constant. The last feature provides a new platform for a `few-body' quantum simulation \cite{cirac:2012,bloch:2012} of the early universe.

The remainder of the paper is organised as follows. Section \S\ref{RelQ} provides a brief overview of the alternative \textit{relational quantization} procedure upon which our treatment of quantum mini-superspace is based. Section \S\ref{Class} presents the main formal details of the classical mini-superspace model. Our initial focus, in Section \S\ref{Tortoise}, is upon developing a coordinate-free representation of the configuration space and its boundary. In Section \S\ref{sing}, we then consider the nature of the classical singularity and introduce the `tortoise' coordinate chart that provides a conformal completion of the configuration space. Finally, in Section \S\ref{sub:explicit_solution}, we give an explicit analysis of the general solutions in order to both isolate their pathological features and offer suggestions as to why one might expect such features to persist in the quantum treatment. The considerable care taken in setting up the classical analysis will reap benefits in the quantum theory. In particular, our classical analysis will be instrumental for isolating and resolving the various quantum pathologies that are encountered. 

We start, in Section \S\ref{sec:observables}, by solving the problem of defining an algebra of self-adjoint quantum observables.  Then, as detailed in Section \S\ref{sec:hamiltonian}, we construct a self-adjoint `Wheeler--DeWitt' Hamiltonian operator and analyse three families of eigenfunctions that are distinguished based upon the value of the cosmological constant. Particularly noteworthy is the  $\Lambda < 0$ family which has a mathematical form that mirrors that of bound Efimov states \cite{efimov:1970} found in few-body quantum physics. Proceeding beyond the standard Dirac analysis, in Section \S\ref{SE for U}, we then apply the procedure of relational quantization leading to the general solution for the quantum cosmological model. An interpretation of the general solution in terms of an experimentally realisable analogue model is then provided in Section \S\ref{analogue}. Finally, in Section \S\ref{sec:sing res}, we provide analysis and discussion of the meaning of cosmic singularity resolution. 

The reader should note that in this paper we will focus on formal analysis of the model and avoid discussion of the interpretational difficulties associated with superpositions of observables in quantum cosmology. A companion paper \cite{Gryb:2017b} provide detailed interpretation and analysis of particular cosmological solutions.

\section{Relational Quantization}
\label{RelQ}

\hspace{11pt} Our approach is based upon an alternative prescription for the quantization of globally reparametrization invariant models. Whereas the standard Dirac quantization approach leads to to a `frozen' Wheeler--DeWitt-type formalism with the accompanying problem of time \cite{Isham:1992,Kuchar:1991,anderson:2012}, our `relational quantization' approach leads to a dynamical quantum formalism with unitary evolution of the universal wavefunction. Key to this formalism is a relational interpretation of the observables where, in the Heisenberg picture, observables operators can evolve according to an unobservable time label whose role in the formalism is to distinguish successive states of the universe. Our approach thus differs from the standard \textit{internal time}-type approaches (variously described as `evolving constants of the motion' or `complete observables') to systems where the classical Hamiltonian is constrained to be zero \cite{Page:1983,Rovelli:1990,Rovelli:1991,Rovelli:2002,gambini:2001,Dittrich:2007,Dittrich:2006,gambini:2009}. In all such approaches, the observables evolve according to a (non-unique) time-dependent Hamiltonian on the physical Hilbert space as dictated by the choice of an internal clock parameter. Such evolution cannot be guaranteed to be unitary and, in fact, non-unitary quantum dynamics obtains for even simple implementations of the internal-time scheme \cite{gambini:2001}. In the relational quantization approach detailed below, evolution is guaranteed to be unitary and the physical Hilbert space can always be unambiguously defined. These two formal advantages of our approach lead to the principal result of this paper.\footnote{We defer detailed discussion of the basis for these advantages to \S\ref{sec:sing res}.}  Our proposal allows for a well-defined quantum formalism that simultaneously meets various conditions for quantum singularity avoidance as defined in the literature. Most significantly, from a physical perspective, we find that the expectation values of all observables in our theory are guaranteed to remain unproblematic even when their classical counterparts break down.

In previous work, relational quantization has been motivated, in general, via independent analyses of the Faddeev--Poppov path integral \cite{gryb:2011}, constrained Hamiltonian methods \cite{gryb:2014} and the Hamilton--Jacobi \cite{Gryb:2016a} formalism of globally reparametrization invariant theories (the final of these will be outlined shortly). All three quantizations rely on a basic observation that the integral curves of the vector field generated by the Hamiltonian constraint in globally reparametrization invariant theories \textit{should not} be understood as representing equivalence classes of physically indistinguishable states since the standard Dirac analysis does not apply to these models \cite{Barbour:2008,Pons:2005,Pons:2010}. On our view, successive points along a particular integral curve should be taken to represent physically distinct moments in time \cite{gryb:2011,gryb:2014,Gryb:2016a} and the vanishing of Hamiltonian does not mean that time evolution must be, rather paradoxically, classified as the `unfolding of a gauge transformation' \cite{Pons:2005rz}.

Relational quantization is based upon the exploitation of an ambiguity in theories that are independent of global time parametrization. This ambiguity relates to the the physical interpretation of the variable, $-\Pi$, that is canonically conjugate to the arbitrary parameter, $t$, associated with the Hamilton flow of the Hamiltonian vector field on phase space, $X_H$. For all globally reparametrization invariant theories, the Hamiltonian is a constraint, and this implies that relevant conjugate variable is a constant. However, one is free to interpret this constant as either an integral of motion -- the value of which is determined by initial conditions -- or as a constant of Nature that takes a certain value fundamentally.

In the context of an Hamilton-Jacobi analysis, this `interpretational ambiguity' leads to a `formal ambiguity' in how we treat the variation of Hamilton's principle function, $S$, with the arbitrary parameter, $t$. The constant-of-motion interpretation leads us to use the standard Hamilton--Jacobi equation:
\begin{equation}\label{eq:TDHJ}
	H\lf(q, \diby S q\rt) = \diby S t\,.
\end{equation}
The time independence of $H$, that is crucial for relational quantization, allows for the simple separation Ansatz:
\begin{equation}\label{eq:Ansatz}
	S(q,t) = \Pi t + W(q)\,,
\end{equation}
which leads to the reduced equation:
\begin{equation}\label{eq:TIHJ}
	H\lf(q, \diby W q\rt) = \Pi\,.
\end{equation}
We can then solve the reduced equation and use \eqref{eq:Ansatz} to derive expressions for the flow of all phase space variables along $X_H$ as parametrised by $t$. Since the flow parameter is monotonically increasing, it acts as a book keeping device to encode the ordering of the relative values of the physical variables.  

Under the alternative interpretation, we treat $\Pi$ as a constant of nature. This implies that $S$ is time independent and, thus, that the reduced equation \eqref{eq:TIHJ} is all we have to describe the physics of the system. In practice, this can be done by applying something like the partial and complete observable program \cite{Rovelli:2002,Dittrich:2006,Dittrich:2007} to describe a set of relational observables for the theory defined by \eqref{eq:TIHJ}. In this approach, the relative values of the physical variables are only ordered relative to an internal clock that need not be monotonically increasing. 

A common misconception regarding the difference between the constant-of-motion interpretation and the constant-of-nature interpretation is that they yield a different degree-of-freedom count and therefore cannot represent the same physics. This misconception arises due to subtle web of relationships between the space of solutions, the space of Dirac observables, the space of couplings, and the space of measurable quantities within these different interpretations. If $\Gamma$ is the unconstrained phase space (assuming there are no other symmetries), then in the constant-of-motion interpretation the space of solutions is the Dim$(\Gamma) - 1$ space of integral curves of $H$. On the other hand, in the constant-of-nature interpretation, one looses 2 degrees of freedom from Dim$(\Gamma)$ because of the Hamiltonian constant $H - E = 0$ and its gauge fixing. However, one gains an extra degree of freedom since the value of $E$ must be fixed in order to specify the Hamiltonian constraint itself. Thus, the dimension of the space of independent solutions is identical in each case. The misconception arises because the space of Dirac observables, which is Dim$(\Gamma) - 2$, is often conflated with the space of independent solutions. While coupling constants are not formally Dirac observables, they must nevertheless be specified in some way within a system in order to for the dynamics to be deterministic. Both interpretations therefore require the same number of physical inputs.

While these two treatments are physically indistinguishable classically, they motivate very different prescriptions for quantization and lead to quantum formalisms between which one could in principle empirically differentiate. If $\Pi$ is understood as a constant of Nature, one follows the standard Dirac quantization route and promotes the reduced equation to a Wheeler--DeWitt-like equation:
\begin{equation}\label{eq:WdW}
	\lf[\hat H - \Pi\rt] \Psi = 0\,.
\end{equation}
If $\Pi$ is understood as a integral of motion, it is natural to follow the relational quantization leading from \eqref{eq:TIHJ} to a Schr\"odinger-type unitary evolution equation:
\begin{equation} \label{eq:TDSE}
	\hat H \Psi = i\hbar \diby \Psi t.
\end{equation}

In the quantum theory, the wavefunctions satisfying \eqref{eq:WdW} and \eqref{eq:TDSE} are different rays in Hilbert space. Moreover, the algebras of physical observables are not unitarily equivalent and, in fact, will not even have the same dimension. We can see this most clearly in terms of the partial and complete observables program where there will not be an operator in the physical Hilbert space corresponding to whichever classical variable is chosen as the internal clock. Contrastingly, in the relational quantization of theories with a global time parameter, the full classical phase space is mapped to the quantum observable algebra.\footnote{For a more detailed discussion of observables in relational quantization and the comparison to Dirac quantization, see \cite{gryb:2014,Gryb:2016a}.}

The purpose of the present paper is to implemented relational quantization explicitly in the context of a  cosmological model. Our approach will rely on the reinterpretation of the role of the cosmological constant, $\Lambda$. Whereas, in most standard treatments, $\Lambda$ is understood as a constant of nature, in our approach it is reinterpreted as a constant of motion in accordance with the role played $-\Pi$ in the above discussion. This leads to a quantum cosmological formalism where the wavefunction of the universe can be in superpositions of  eigenstates of $\Lambda$. Our approach is, thus, naturally connected to the unimodular approach to gravity \cite{Unruh:unimodular_grav}. As noted above, a cosmological model of the same form as that studied here was derived in the context of that program \cite{Unruh_Wald:unimodular}. Whilst the unimodular approach to gravity was vulnerable to the objection that the the unimodular condition appears incompatible with foliation invariance \cite{kuchar:1991a}, there is the potential for the application of our proposal to the quantization of gravity when understood in terms of the \emph{shape dynamics} formalism \cite{gryb:shape_dyn,barbour_el_al:physical_dof,gryb:2_plus_1}. This is due to the fact that, within shape dynamics, foliation invariance is not manifestly broken. Rather, shape dynamics re-encodes the gravitational degrees of freedom in terms of conformally invariant three-geometries and then describes their evolution in terms of a global time. Understanding our mini-superspace model as an approximation to quantum shape dynamics rather than unimodular gravity lends plausibility to the idea of a Schr\"{o}dinger equation for the universe.

\section{Classical Mini-Superspace Cosmology}\label{Class}

\subsection{Configuration Space Geometry}\label{Tortoise}

\hspace{11pt} The model we will consider is an homogeneous and isotropic FLRW universe with zero spatial curvature ($k=0$) described by the scale factor, $a$, coupled to a massless free scalar field, $\phi$. In terms of these variables the space-time metric takes the form:
\begin{equation}
	\de s^2 = - N(t)^2 \de t^2 + a(t)^2 \lf( \de x^2 + \de y^2 + \de z^2 \rt)\,
\end{equation}
with $\partial_i \phi = 0$. The symmetry reduced action becomes:
\begin{multline}\label{eq:miniaction}
S=S_\text{EH}+S_\phi = -  \frac 1 {2 \kappa} \int_\mathbbm{R} \de t\, V_0 a^3 \lf[ \frac 6 N \lf( \frac {\dot a} a \rt)^2 + 2 N \Lambda \rt]\, \\ + \int_\mathbbm{R} \de t\, \frac{ V_0 a^3}{2N} \dot \phi^2\,,
\end{multline}
where $\Lambda$ is the cosmological constant, $\kappa = 8\pi G$ and
\begin{equation}
	V_0 = \int_{\Sigma'} \de^3 x \sqrt{g}\,.
\end{equation}
Here, $g_{ab}$ is the intrinsic metric of a space-like Cauchy surface $\Sigma$ and $\Sigma' \subset \Sigma$ is a `large' space-like fuducial cell used to give meaning to relative spatial volumes. All quantities are scalars on $\mathbbm R$, which labels successive instants of time. The Hamiltonian corresponding to \eqref{eq:miniaction} is: 
\begin{equation}\label{eq:miniham}
	H = N \lf[ - \frac \kappa {12 V_0 a} \pi_a^2 + \frac{1}{2V_0 a^3} \pi_\phi^2 + \frac{V_0 a^3}\kappa \Lambda \rt]\,,
\end{equation}
while the symplectic form is such that $\pi_a$ and $\pi_\phi$ are canonically conjugate to $a$ and $\phi$. The Lagrange multiplier $N$ enforces the on-shell vanishing of the Hamiltonian. The Hamiltonian induces a vector field on phase space $\Gamma(a, \pi_a; \phi, \pi_\phi)$ via the canonical symplectic form, and the integral curves of this vector field represent the unparametrized classical solutions of the theory. As noted in \S\ref{RelQ}, we believe that the integral curves in question should not be understood as representing equivalence classes of physically indistinguishable states according to the standard Dirac analysis.

To understand the general features of the classical solutions, and the particularities of the quantum theory, it will prove useful to develop a coordinate-free representation of this model in terms of the geometry of configuration space, $\mathcal C$. Towards this end, we first express our configuration variables in terms of relative spatial volume (in the slicing selected by homogeneity and isotropy), $v$, and the dimensionless value of the scalar field, $\varphi$:
\begin{align}
	v &= \sqrt {\frac 2 3} a^{3} & \varphi &= \sqrt{\frac {3 \kappa}2} \phi\,.
\end{align}
Note that these definitions in terms of dimensionless quantities imply that the dimension of all canonical momenta will be that of angular momenta. This can be absorbed into a single angular momentum scale, $\hbar$, which at the level of the classical theory merely serves as a bookkeeping device. If we conveniently identify this reference scale with the Planck scale, we can use it to identify the regime where the classical formalism is expected to break down.

Using $\hbar$, it is convenient to define the dimensionless lapse, $\tilde N$, and cosmological constant, $\tilde \Lambda$ as
\begin{align}
	\tilde N &=  \sqrt{ \frac 3 2} \frac {\kappa \hbar^2 v N}{V_0}  & \tilde \Lambda &= \frac{ V_0^2} {\kappa^2\hbar^2} \Lambda \,,
\end{align}
so that the Hamiltonian, in this coordinate chart, becomes
\begin{equation}\label{eq:v-phi Ham}
	H = \tilde N \lf[ \frac{1} {2\hbar^2} \lf( - \pi_v^2 + \frac{\pi_\varphi^2}{v^2} \rt) + \tilde \Lambda \rt]\,,
\end{equation}
where $\pi_v = \sqrt{\frac 1 6} a^{-2}\pi_a $ and $\pi_\varphi = \sqrt{\frac 2 {3 \kappa}} \pi_\phi $ are the conjugate momenta to $v$ and $\varphi$. This representation of the Hamiltonian generates evolution parametrized by the rescaled time $\tilde t =\sqrt { \frac 2 3 }  \frac {V_0}{\kappa\hbar^2 v} t$, which is  proportional to standard cosmological time corresponding to $N = 1$.

After making these identifications, we find that the natural metric on $\mathcal C$ relevant to the dynamical problem in consideration is the flat Minkowski metric in 2d:
\begin{equation}
	\de s^2_\mathcal{C} = \eta_{AB} q^A q^B\,,
\end{equation}
where $(A,B) = 1 \text{ and } 2$, the $q^A$ represent some arbitrary coordinate chart on $\mathcal C$, and $\de s_\mathcal{C}$ is the dynamically relevant proper distance element on $\mathcal C$. That this is indeed the appropriate metric can be seen from the fact that the action \eqref{eq:miniaction} takes the form
\begin{equation}\label{eq:gen action}
	S = \int \de t \lf[ \frac 1 {2{\tilde N}} \eta_{AB} \dot q^A \dot q^B + \tilde N \tilde \Lambda \rt]\,,
\end{equation}
where $q^A = (v, \varphi)$ and the Minkowski metric is expressed in a hyperbolic polar coordinate chart where $\eta_{AB} = \hbar^2\text{diag}(-1, v^2)$. In terms of this metric, the Hamiltonian constraint takes the form
\begin{equation}\label{eq:class H}
	H = \tilde N \lf[ \frac 1 2 \eta^{AB} p_A p_B + \tilde \Lambda \rt]\,,
\end{equation}
which states that the generalized momenta, $p_A$, are fixed-norm vectors on $\mathcal C$.

Geometrically, $\mathcal C$ is not the entire Minkowski plane because $v$ is restricted to $\mathbbm R^+$. Rather, $\mathcal C$ can be identified with the upper Rindler wedge, which is defined to be such that the proper distance between the origin and arbitrary points in the wedge is positive. This space has a boundary, $\partial \mathcal C$, on the light-cone centred on the origin where $v=0$, and this boundary leads to the most physically important properties, both classical and quantum, of this cosmological model. The geometry of Rindler space is well-known (see, for example, \cite{Socolovsky:2013rga} for a review). We will now review some key features that will be useful in our construction.

The boundary of Rindler can be identified in a coordinate-free way by using the global Killing vector fields of $\eta_{AB}$. Because $\eta_{AB}$ is the flat Minkowski metric in 2D, its Killing vectors form an ISO(1,1) algebra that can be parametrized in terms of two translation generators and one boost generator. The boundary is the union of the one dimensional surfaces spanned by the two translation generators where the boost generator becomes asymptotically null. Inspired by the form of the metric in $(v,\varphi)$ coordinates, we can identify the boost parameter with $\varphi$. Our definition of $\partial \mathcal C$ is then clearly only compatible with the ISO(1,1) algebra when $\varphi \to \pm \infty$, which is consistent with the claim above that $\partial \mathcal C$ is the null cone centered at the origin. The boost generator, which is hypersurface orthogonal to surfaces of constant $\varphi$, and the unique vector field orthogonal to it, which is hypersurface orthogonal to surfaces of constant $v$, provide a geometrically privileged set of coordinates -- up to reparametrization and orientation -- consisting of $\varphi$ and $v$. Note that, while the integral curves of the translation generators end on the part of the boundary that is generated by the orthogonal translation, the integral curves of the boost generators can be continued indefinitely. The boost, therefore, represents the only symmetry of $\mathcal C$ valid everywhere in the space, and will be useful for conveniently splitting the solution space in the quantum theory.

\subsection{Classical Singular Behaviour}\label{sing}

The existence of classical singularities can be understood in terms of the incompleteness of causal curves and the existence of some type of curvature pathology \cite{penrose:1965,hawking:1970,hawking:1973,thorpe:1977,ellis:1977,senovilla:1998,curiel:1999}. The physical significance of such features can be made apparent by considering the phenomenology of a test particle traveling into a singular region. While the incompleteness of a causal curve corresponds to the test particle traversing the complete extension of the curve in finite proper time, a curvature pathology corresponds to the particle experiencing unbounded tidal forces. Generic statements about the precise character of spacetime singularities are difficult to come by, not least since curvature pathologies can be shown to be neither necessary nor sufficient for curve incompleteness \cite{ellis:1977}. For our purposes, it will be sufficient to follow the logic of the  Penrose-Hawking singularity theorems \cite{penrose:1965,hawking:1970,hawking:1973} and consider the behaviour of the expansion parameter of a congruence of geodesics. That the expansion parameter becomes unbounded and negative somewhere along the congruence in a finite proper time is taken as an indication that the relevant spacetime contains incomplete causal curves. Additionally, we will take curvature pathologies to be signalled by a divergence in any Kretschmann invariant. A spacetime region containing both such features is unquestionably a singular one. In \S\ref{sub:explicit_solution} we will demonstrate explicitly that our model is singular in both the curve incompleteness and curvature pathology sense.\footnote{The singular nature of the FLRW model is treated explicitly in \cite{senovilla:1998} with $\Lambda = 0$. Here we are principally interested in the $\Lambda > 0$ case. For this class of models the singularity theorems do not, in fact, apply since the relevant energy conditions do not hold. For $k=0$ we nevertheless recover singular behaviour for generic classical solutions in the senses described above.}

Classically, the action \eqref{eq:gen action} enforces a geodesic principle on $\mathcal C$ in terms of the metric $\eta$.\footnote{One way to see this is to integrate out $\tilde N$ by evaluating the action using the equations of motion for $\tilde N$. This shows that the action is proportional to the geodesic length of a curve on $\mathcal C$ using the metric $\eta$.} The classical solutions are geodesics of 2d Minkowski space: straight lines. By inspecting the Hamiltonian constraint and momentarily undoing the Legendre transformation, it is straightforward to see that $\Lambda$ controls the sign (and magnitude) of infinitesimal proper-distance along classical curves in $\mathcal C$. Thus, `time-like' geodesics correspond to $\Lambda >0$, `null' geodesics correspond to $\Lambda = 0$, and `space-like' geodesics correspond to $\Lambda < 0$. 

\begin{figure}[h]
	\includegraphics[width = \linewidth]{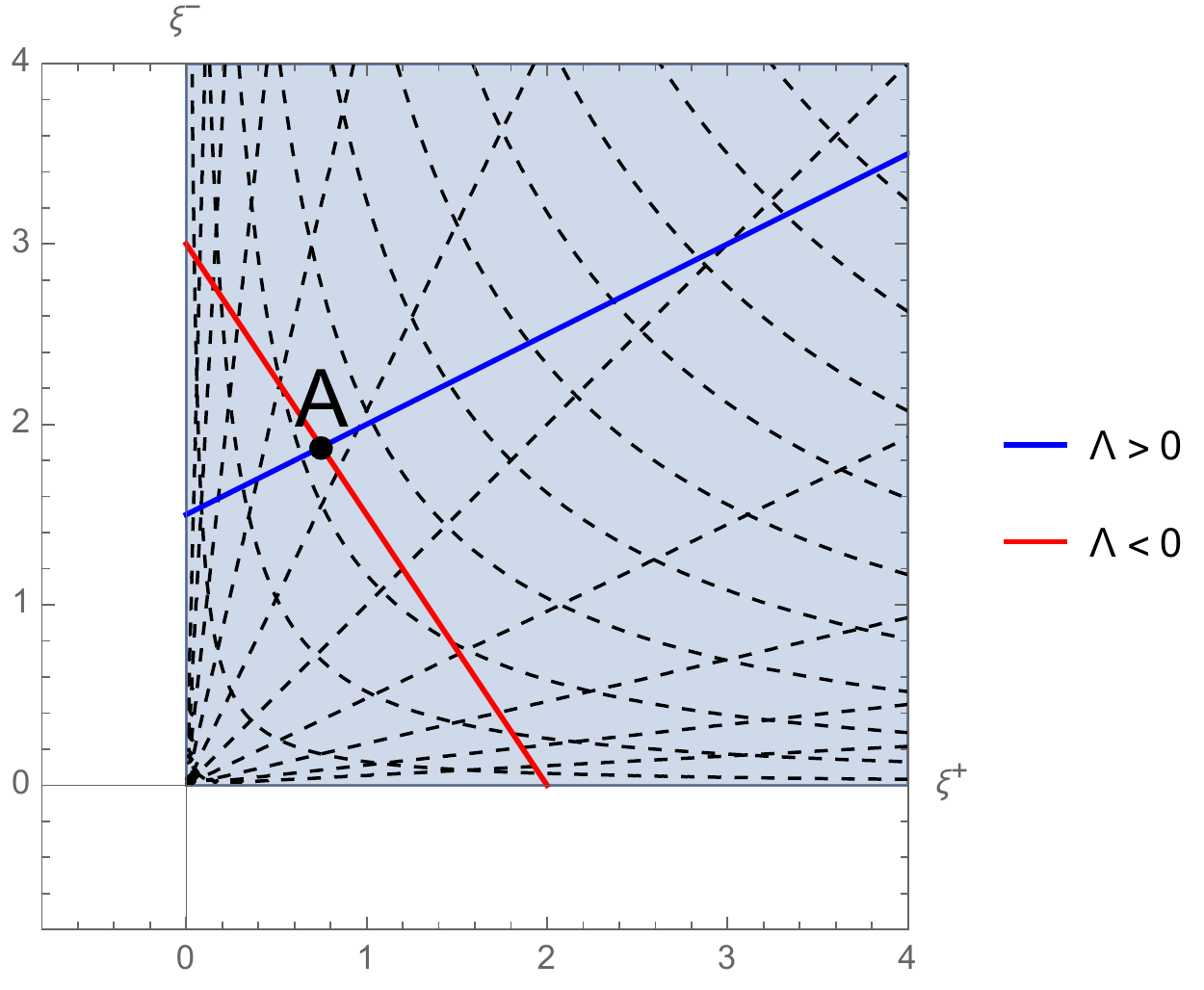}
	\caption{\label{fig:class_solns}Two examples of typical classical solutions with positive (in blue) and negative (in red) cosmological constant as expressed in light-cone coordinates $\xi^\pm = x^0 \pm x^1$, where $x^A$ are standard Cartesian coordinates in the upper Rindler wedge. Lines of constant $a$ (the hyperbolas) and $\phi$ (the straight lines) are drawn in dashed lines. The solutions live in the restricted domain indicated by the shaded region. It is clear that all solutions will cross the singular boundary, $\partial \mathcal{C}$, at least once. The units are arbitrary.}
\end{figure}

The presence of the boundary $\partial \mathcal{C}$ implies that $\mathcal C$ is geodesically incomplete -- unlike the full Minkowski plane. This can be deduced from the fact that, as mentioned at the end of \S\ref{Tortoise}, the translational Killing vectors are orthogonal to the boundary -- but it is perhaps most easily seen by considering the space of solutions in a simple coordinate chart, say standard light-cone coordinates on $\mathbbm R^{2+}$. In this chart, $\mathcal C$ is the upper-right quadrant and the geodesics are straight lines where the sign of the slope is proportional to the sign of the cosmological constant. As illustrated in Figure~\ref{fig:class_solns}, any geodesic passing through an arbitrary point $A$ on $\mathcal C$ that is not on the boundary will cross the boundary at least once. This happens exactly once for non-negative $\Lambda$ and twice for negative $\Lambda$. Because the metric is finite along the entire curve in this chart, the proper-distance will also be finite. This means that all geodesics passing through $A$ will terminate on the boundary in finite proper distance, thus proving geodesic incompleteness (of $\mathcal C$ \emph{not} spacetime).

While the set $(\mathcal C, \eta)$ itself is geodesically incomplete, it is possible to construct a conformal completion $(\mathcal C_0, \eta_0)$ that is geodesically complete and where $\mathcal C_0$ is globally isomorphic to $\mathcal C$. This can be achieved by defining the reference metric
\begin{equation}
	\eta^0_{AB} = \Omega^2 \eta_{AB}\,,
\end{equation}
where, in the $(v, \varphi)$ chart, $\Omega = 1/v$. If we define the \emph{tortoise} coordinate $\mu$ such that
\begin{equation}
	\mu = \log v\,,
\end{equation}
which respects the split suggested by the boost symmetry of $\mathcal C$, then it is easy to see that, in the $(\mu, \varphi)$ coordinates, $\eta^0_{AB} = \text{diag}(-1,1)$, which is the Minkowski metric on the full $\mathbbm R^{(1,1)}$ plane. This conformal completion and the associated tortoise coordinate will be essential tools in defining the eigenfunctions of the momentum operator in the quantum theory \S\ref{sec:observables}.

It is important to distinguish between the geodesic incompleteness of $\mathcal C$ and the geodesic incompleteness of the \emph{space-time} metric $g_{\mu\nu}$. Although the former may be sufficient for the latter, the two senses of incompleteness are logically distinct. In the context of the minisuperspace model, we will see that the boundary $\partial \mathcal C$  corresponds to the  configuration space point where we have both that: i) the expansion parameter of some congruence becomes negative and unbounded, implying that the \emph{space-time} geodesics terminate in finite proper time; and ii) there is a curvature pathology. This can be seen straightforwardly in our chosen coordinate chart: the condition $v = 0$ both defines the boundary of $\mathcal C$ and can be connected to a singularity in the senses i) and ii) above. This will be shown explicitly in \S\ref{sub:explicit_solution} below.  All non-trivial cosmological solutions of our model thus contain at least one \emph{big bang} or \emph{big crunch} singularity, and these occur precisely at the boundary $\partial \mathcal{C}$.

\subsection{General Explicit Solutions} 
\label{sub:explicit_solution}

The geometric methods used in the previous sections, though valuable for highlighting generic features of the solution space, do not provide the most physically enlightening tools for investigation of certain features that will prove fundamental to the quantum analysis. Rather, to both connect our solutions to  cosmological observations and to construct the quantum formalism, it is valuable to describe general solutions in the observationally relevant coordinate chart where the configuration variables are $v$ and $\varphi$.\footnote{As we will see in \S\ref{sec:observables}, quantization is \emph{not} chart dependent.} In this section, we will restrict to $\Lambda > 0$ since, as will be shown explicitly in \S\ref{sec:hamiltonian}, only these cases will have a well-defined and non-trivial semi-classical limit.

The classical analysis in these variables can be performed by integrating Hamilton's equations generated by the Hamiltonian \eqref{eq:v-phi Ham}. Hamilton's second equation for $\varphi$ tells us that $\pi_\varphi$ is a conserved quantity, which we will call $k_0$ in anticipation of the variables used to construct our cosmological quantum state. Hamilton's first equation for $\dot v$ can then be inverted and inserted into to the Hamiltonian constraint leading to 
\begin{equation}
	\lf( \frac {\de \tau}{\hbar} \rt)^2 = \frac{ \de v^2 }{ 2 \tilde\Lambda + \frac 1 {v^2} \lf(\frac{k_0}{\hbar}\rt)^2 }\,,
\end{equation}
where $\de \tau = \tilde N \de \tilde t$ (this is precisely Friedmann's first equation for this system). Integration of this equation is straightforward, although care must be taken to interpret different branches of the solution in terms of expanding and/or contracting phases, which differ by a choice of the arrow of time. Solutions with a contracting `crunch' phase followed by an expanding `bang' phase are obtained by the branch whose integral of motion is given by
\begin{equation}
	v(\tau)^2 = - \frac{k_0^2}{\omega_0^2} + \frac{\omega_0^2}{\hbar^4} \tau^2\,,
\end{equation}
where we have defined the quantity
\begin{equation}
	\omega_0 \equiv \sqrt{2 \hbar^2 \tilde \Lambda}\,,
\end{equation}
which will be useful later in our treatment of the quantum theory. In this solution, we have used the time translational invariance of the theory so that the classically singular points of the solution, where $v = 0$, occur symmetrically about $\tau = 0$ at the singular times
\begin{equation}
	\pm \tau_\text{sing} \equiv \pm \frac {\hbar^2 k_0}{\omega_0^2}   \,,
\end{equation}
which represent the classical Big Bang and Crunch.

That these times represent a classical singularity in the sense defined in \S\ref{sing} above can be seen as follows. Consider the time-like congruence that is hypersurface orthogonal to the homogeneous and isotropic spatial geometries that define our constant-$\tau$ slices. By homogeneity, the expansion parameter, $\theta$, of this congruence is constant along these slices. Because these slices are the same as those used in our canonical split, the value of $\theta$ is equal to the trace of the extrinsic curvature tensor, $g^{ab} K_{ab}$, of our canonical slices. A straightforward canonical analysis then indicates that
\begin{equation}\label{singdef}
	\theta \propto -\pi_v/ \tilde N\,.
\end{equation}
The form of the Hamiltonian constraint,
\begin{equation}
	\pi^2_v = \frac{\pi^2_\varphi}{v^2} - 2 \hbar^2 \tilde \Lambda\,,
\end{equation}
implies that $|\pi_v| \to \infty$ as $v \to 0$ since $\pi_\varphi$ is a constant of motion. This implies that $\theta \to -\infty$ for the branch of solution we are interested in. The condition $k = 0$ means that the 3d Ricci scalar vanishes. Given this, the divergence of the trace of $K_{ab}$ indicates (via the Gauss-Codazzi relations) that the 4d Ricci scalar is also divergent. This confirms that, when $v = 0$, we have both incompleteness of causal curves, in the sense of the Penrose-Hawking singularity theorems, and a curvature pathology, in the sense that the Ricci scalar diverges.
 
\begin{figure}[h]
	\subfloat[Spatial volume $v(\tau)$.]{
		\includegraphics[width=\linewidth]{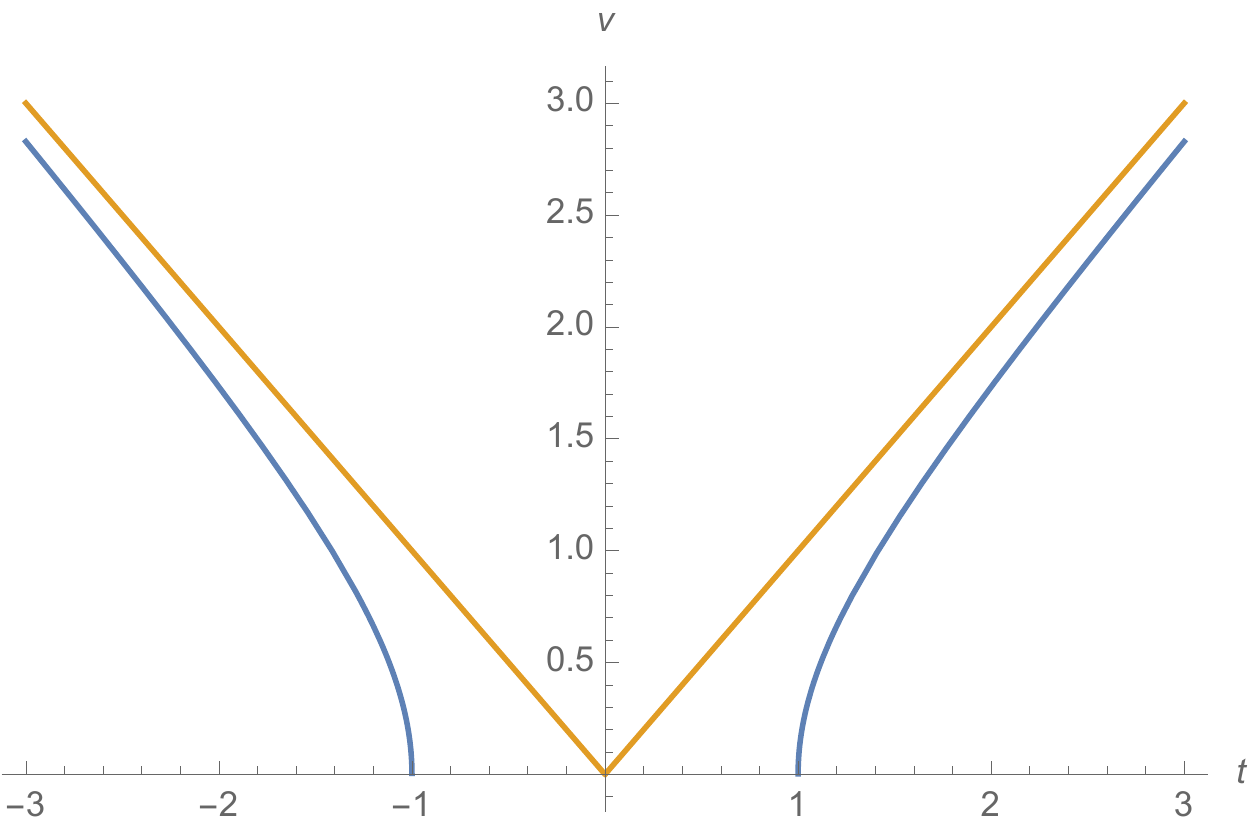}
		\label{fig:xplot}
	}\\
	\subfloat[Scalar field, $\varphi(\tau)$.]{
		\includegraphics[width=\linewidth]{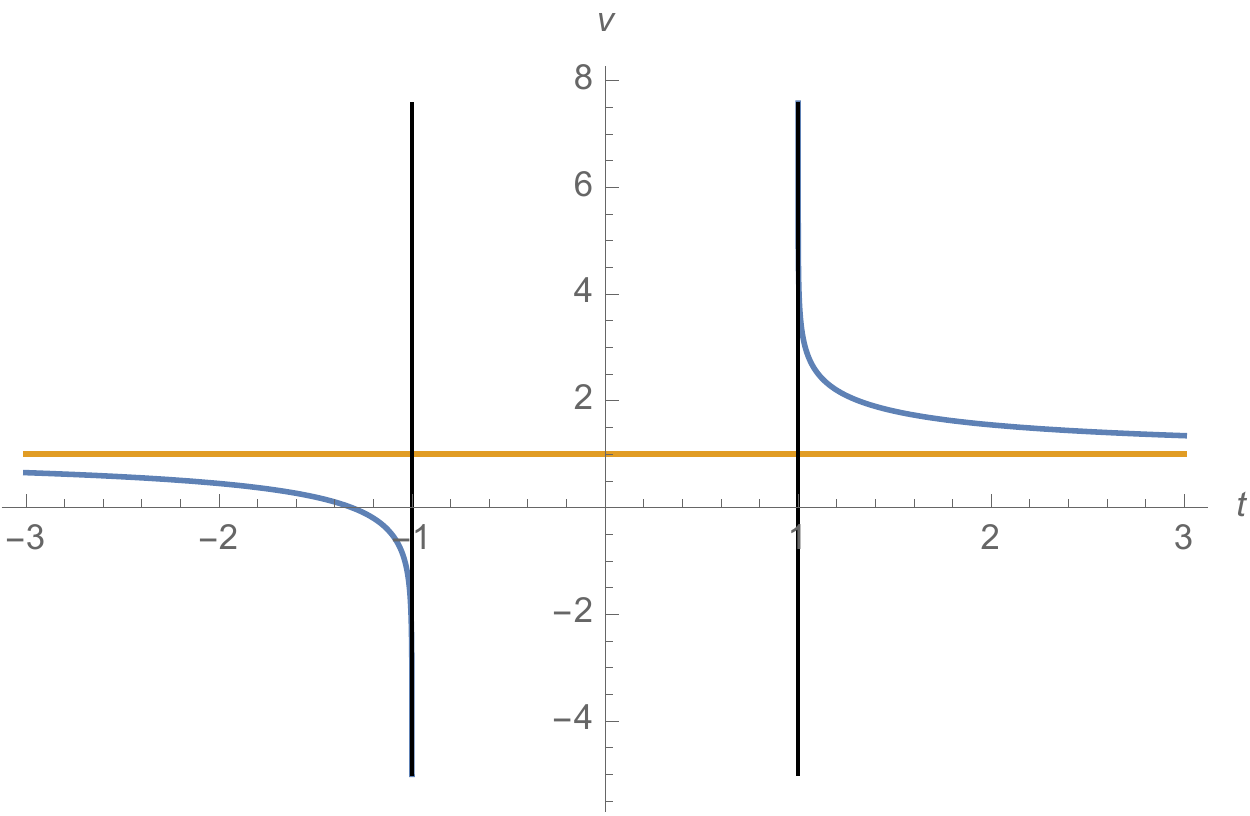}
		\label{fig:nuplot}
	}
	\caption{\label{fig:class solns}Classical solutions for small $\varphi_\infty$. Asymptotic values of the parameters are shown in yellow. Singular lines drawn in black. Temporal units are chosen so that $\tau_\text{sing} = 1$. Units for $v$ are chosen so that $s =1$.}
\end{figure}

The integration constants $k_0$ and $\omega_0$ can be used to define a dimensionless parameter
\begin{equation}
	s \equiv \abs{\frac {k_0} {\omega_0}}
\end{equation}
that sets the (relative) scale where the dynamics of $\varphi$ become relevant to the evolution of $v$. At late and early times, $\tau/\tau_\text{sing} \gg 1$, the dynamics of $v$ become indifferent to the value of $s$, and the system enters an approximate de~Sitter expansion phase:
\begin{equation}
	x(\tau) \to \frac{\omega_0 \abs{\tau}}{\hbar^2} = \sqrt{2\tilde \Lambda} \frac {\abs{\tau}} \hbar\,.
\end{equation}
Figure~\ref{fig:xplot} illustrates the general behaviour of this solution.

It is now possible to integrate Hamilton's first equation for $\varphi(\tau)$. Care must again be taken to set limits of integration such that the resulting integration constants can be straightforwardly interpreted. A convenient choice leads to
\begin{equation}
	\varphi(\tau) = \varphi_\infty + \tanh^{-1}\lf( \frac {\tau_\text{sing}} {\tau} \rt)\,,
\end{equation}
where we interpret $\varphi_\infty$ as the late and early asymptotic value of $\varphi$. The divergence of $\varphi(\tau)$ at $\tau = \tau_\text{sing}$ represents a genuine physical divergence and is consistent with the singular behaviour observed above. Figure~\ref{fig:nuplot} illustrates the general behaviour. The translational invariance of $\eta_{ab}$ in $\varphi$ (which are boosts in $\mathcal C$) implies that the solutions themselves have symmetries under shifts in $\varphi$.

\begin{figure}[h]
	\centering
	\includegraphics[width=\linewidth]{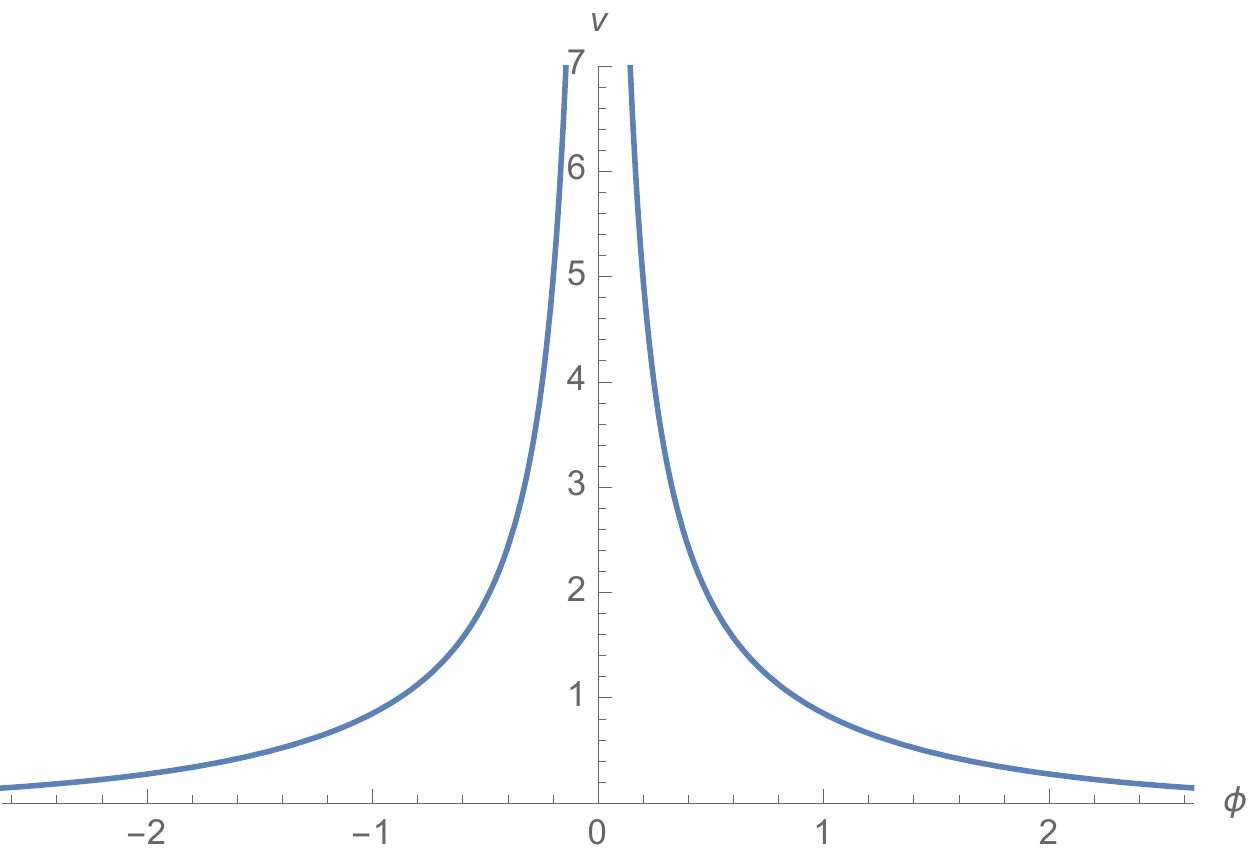}
	\caption{Classical solutions on $\mathcal C$ in units where $s = 1$ and $\varphi_\infty = 0$.} \label{fig:rep inv sols}
\end{figure}

Reparametrization invariant curves on $\mathcal C$ can be obtained by parametrizing the solution for $v$ in terms of $\varphi$ by eliminating $\tau$. This is possible because $\varphi$ is monotonic along our chosen branch and represents, therefore, a good clock. Branches that contract and then re-expand in this way can be conveniently written as
\begin{equation}
	v = s \abs{\text{cosech}(\varphi-\varphi_\infty)}\,.
\end{equation}
Figure~\eqref{fig:rep inv sols} illustrate these curves on $\mathcal C$.
Note that the solution illustrated in Figure~\eqref{fig:rep inv sols} characterizes \emph{all} physically meaningful solutions to the system. This is because the arbitrary units of space and time can be used to arbitrarily rescale the two constants of motion $\Lambda$ and $k_0$. What is left is the un-parametrized dimensionless curve of Figure~\eqref{fig:rep inv sols}. We will see that the time-dependent wavefunction that can be defined over-top of this space can have a meaningful size in relation to the classical solution. This implies that an additional parameter -- representing some dimensionless version of $\hbar$ -- labels the relative size of quantum effects.

It is instructive to consider at this point a very general argument as to why we might expect that the singularity should \textit{not} be resolved in the quantum theory. Consider the generalised Ehrenfest theorem
\begin{equation}\label{Ehrenfestpi}
\frac{\partial}{\partial t} \Big{<}\hat{\pi}_v(t)\Big{>} =\frac{1}{i\hbar} \Big{<}[\hat{\pi}_v(t),\hat{H}]\Big{>} + \Big{<}  \frac{\partial \hat{\pi}_v (t)}{\partial t}\Big{>}\,
\end{equation} 
under the observation that $\theta \propto -\pi_v/ \tilde N$. Given that we have just shown that the classical value of $\pi_v$ is problematic, as indicated by the fact that $v \to 0$, one might expect Ehrenfest's theorem to imply that the quantum expectations values should be equally problematic. More precisely, the logic of the Hawking--Penrose singularity theorem is to give very general conditions under which the time derivative of the expansion parameter, $\theta$, of some congruence of geodesics in the spacetime is always negative, and that the expansion parameter itself will eventually go to minus infinity. What we have shown in this section is that the conditions for this theorem are met, in the case of minisuperspace models with vanishing spatial curvature, even when the cosmological constant is positive so that the strong energy condition is violated. It follows that the same logic can be straightforwardly applied to argue that, given the Ehrenfest theorem and the requirement that the classical Poisson algebra should be faithfully represented by the quantum operator algebra, the expectation value of $\dot \theta$ should be always negative, and, therefore, $\mean{\theta} \to -\infty$. Thus, provided that the conditions for the relation \eqref{Ehrenfestpi} hold, one should expect a kind of quantum generalisation of the classical singularity theorem. Fortunately, as we will see in \S\ref{sec:observables} below, the $\hat \pi_v$ operator fails to be self-adjoint, meaning that the conditions for \eqref{Ehrenfestpi} are violated and a quantum version of the singularity theorem can be avoided.


\section{Quantum Mini-Superspace Cosmology}
\subsection{Algebra of Observables} \label{sec:observables}

In the quantum theory, the existence of the classically singular boundary $\partial \mathcal C$ leads to difficulties in finding self-adjoint representations of symmetric operators in Hilbert space.\footnote{See \cite{horowitz:1995} for a discussion of the connection between self-adjointness and singular spacetimes.} Specifically, the presence of a boundary allows for the possibility of boundary terms to contribute to the action of the adjoint, and self-adjointness requires that these boundary terms vanish. For a particular representation of some operator, three distinct possibilities arise:
\begin{enumerate}[i)]
	\item All states, which are square integrable on the appropriate interval and measure, automatically satisfy the conditions required for the boundary terms to vanish. In this case, the operator is termed \emph{essentially self-adjoint}.
	\item All square integrable functions are spanned by the solutions of a unitary family of boundary conditions that guarantee the vanishing of the boundary terms. This implies an underdetermination in the self-adjoint representations of the operator in question, and the solution space of each member of the unitary family defines the domain of a particular \emph{self-adjoint extension} of the operator.
	\item There exist square integrable functions such that the boundary terms are non-zero. In this case, the operator is not self-adjoint or, rather, is said to have no self-adjoint extensions.
\end{enumerate}
A theorem, due to von~Neumann, allows us to diagnose into which of the three categories a particular symmetric operator falls  based upon the dimensionality of the relevant `deficiency subspaces' (see \cite{reed:1975} theorem X.1). A further theorem implies that all real symmetric operators fall in to category i or ii  (see \cite{reed:1975} theorem X.3). In our analysis, we will focus on the behaviour of the relevant boundary terms, which have a more direct geometric significance, to diagnose the self-adjointness of our operator algebra. This method can be shown to be equivalent to the von~Neumann treatment \cite{gitman:2012}. 

From now on, we will consider building representations of our quantum operator algebra using point-wise multiplication operators of the coordinates in a particular chart of $\mathcal C$ and shift operators of these coordinates in that chart (see the definitions \eqref{eq:operator defn} below). This represents an infinite family of conjugate operators corresponding to classical phase space functions, where each chart provides a pair of operators. Changes of coordinates on $\mathcal C$ induce transformations between pairs of operators.\footnote{Note that this is different from considering different representations of the \emph{same} operators in different charts.} Na\"ively, one would expect that these operators should be unitarily equivalent since they correspond to classically equivalent variables. However, there are obstructions to defining the self-adjoint shift operators of the coordinates of an arbitrary chart of $\mathcal C$. A detailed analysis of these issues in terms of the formal structure of quantization has been given by Isham \cite{isham:1984}. For our purposes, we will see these obstructions arise through the geometric properties of the states and operators. In particular, there is a mismatch between the condition for square integrability on states, which transforms like a scalar on $\mathcal C$, and the condition for the self-adjointness of the momentum operators, which transforms like a co-vector. This mismatch implies that the standard canonical quantization procedure leading to an algebra of self-adjoint operators is not well-defined in all charts over $T^* \mathcal C$.

It is important to note that these obstructions are compatible with the Stone--von~Neumann theorem, which assumes that the classical phase space is $\mathbbm R^{2N}$, with $N$ the number of Lagrangian degrees of freedom. This assumption is violated in the problematic charts of $\mathcal C$. For example, the physically natural variable $v$ has the phase space $\mathbbm R^+ \times \mathbbm R$, and will be seen to be problematic below. In particular, we will see that its momentum operator, $\hat \pi_v$, will not be self-adjoint. As discussed at the end of \S\ref{sub:explicit_solution} above, rather than being problematic, this obstruction is important for evading a quantum generalisation of the singularity theorem, and is ultimately responsible for allowing singularity resolution.

In order for the quantum observable algebras to reduce to the classical one in the appropriate limit, it is still necessary to find a well-defined momentum operator to replace $\hat \pi_v$. Fortunately, Darbaux's theorem guarantees that there exists a coordinate chart where $T^* \mathcal C = \mathbbm R^{2N}$ in some finite patch. Moreover, in our model, it is possible to find a global set of coordinates that has this property, and this set of coordinates will provide us with a well-defined momentum operator. In the semi-classical limit, the eigenvalues of this operator can be used to reconstruct the classical observable algebra in all charts of $T^* \mathcal C$.

We will now give the details of an explicit geometric construction of representations of the operator algebra of our quantum theory. Given the geometric considerations of the previous section, we consider the Hilbert space,  L$^2(\mathcal C, \de^2 q \sqrt{-\eta})$, of square integrable functions on $\mathcal C$ under the Borel measure $\de \theta = \de^2 q \sqrt{-\eta}$, where $\eta = \det \eta_{AB}$. This space includes all complex functions $(\Phi,\Psi)$ that obey 
\begin{equation}\label{eq:inner prod}
	\mean{\Phi, \Psi} \equiv \int_\mathcal{C} \de^2 q \sqrt{- \eta}\, \Phi^\dag \Psi < \infty \,.
\end{equation}
To convert this into a condition on states, we make use of the conformal completion $(\mathcal C^0, \eta^0)$ of $(\mathcal C, \eta)$ discussed in \S\ref{sing}. In terms of the tortoise coordinate $\mu$ and $\varphi$, a well-known result, \cite[Lemma 2.13]{gitman:2012}, establishes that the condition
\begin{equation}\label{eq:inner prod ref}
	\mean{\Phi, \Psi} \equiv \int_{\mathbbm R^2} \de \mu \de \varphi \sqrt{- \eta^0}\, \Phi^\dag \Psi < \infty\,,
\end{equation}
combined with the requirement that the momentum eigenstates are also in the Hilbert space, implies
\begin{align}\label{eq:si req}
	\Phi^\dag \Psi(\pm \infty,\varphi) &\to 0 & \Phi^\dag \Psi(\mu,\pm \infty) &\to 0  \,.
\end{align}
The falloff conditions above can be adopted to an arbitrary measure by multiplying and dividing the integrand of \eqref{eq:inner prod ref} by $\sqrt{-\eta_0}$. The condition \eqref{eq:si req} then becomes\footnote{Note that this follows because the limits $\mu \to \pm\infty$  and $\varphi \to \pm \infty$ correspond to the boundary $\partial \mathcal C$.}
\begin{equation}\label{eq:si gen}
	\lf. \sqrt{\frac{\eta}{\eta_0}}\, \Phi^\dag \Psi \rt|_{\partial \mathcal{C}} \to 0 \,,
\end{equation}
which, for an arbitrary state, converts to the necessary condition:
\begin{equation}
	\lf. \lf( \frac \eta {\eta_0} \rt)^{1/4} |\Psi | \rt|_{\partial \mathcal C} \to 0\,
\end{equation}
for square integrability. This scalar condition on the states is not compatible with the condition for the self-adjointness of all momentum operators.

To see this explicitly, consider the following representation of the infinite family of point-wise multiplication and shift operators discussed above:
\begin{align}\label{eq:operator defn}
	\hat q^A \Psi &= q^A \Psi & \hat p_A \Psi &= -i\hbar (-\eta)^{-1/4}\diby{}{q^A}\lf[ (-\eta)^{1/4}\Psi\rt]\,.
\end{align}
The ordering is chosen to be symmetric under the chosen measure. Integration by parts tells us that
\begin{multline}\label{eq:p_A boundary}
	\mean{\Phi, \hat p_A \Psi} = \mean{\hat p_A \Phi, \Psi} - \oint_{\partial\mathcal C} n_A \de l \sqrt{\chi} \Phi^\dag \Psi\,,
\end{multline}
where $\de l$ is an integration measure on $\partial \mathcal C$, $\sqrt{\chi}$ is the pullback of $\sqrt{-\eta}$ onto $\partial \mathcal C$ and $n_A$ is a unit normal to the boundary. Because $\sqrt{\chi}$ splits into a volume form on $\partial \mathcal C$ and a 1-form in $\mathcal C$, the condition
\begin{equation}\label{eq:p s.a. ness}
	\lf.  n_A \de l \sqrt{\chi}\, \Phi^\dag \Psi \rt|_{\partial \mathcal C } \to 0\,,
\end{equation}
transforms, as expected, as a co-vector and is, therefore, chart-dependent. This chart dependence implies that not all square integrable functions will satisfy the condition of self-adjointness for every choice of coordinates.

Any canonical quantization procedure leading to an algebra of self-adjoint operators will, therefore, break the elegant coordinate-free construction presented thus far. It will then be necessary below to consider specific coordinate representations of our operators. Note that the same is not true for scalar operators, whose boundary term will transform in the same way as the square integrability condition. The Hamiltonian operator presented below in \S\ref{sec:hamiltonian} (see equation~\eqref{eq:gen Ham}) is an example of such a scalar function and exists in all possible coordinate representations.

A particularly convenient coordinate split can be achieved by taking advantage of the global Killing vector field provided by the boosts. As discussed in \S\ref{Tortoise}, the boost symmetry of $\eta_{AB}$ suggests a coordinatization of $\mathcal C$, up to reparametrization, in terms of the boost parameter $\varphi$ and the orthogonal coordinate $v$. The $\varphi$-dependent part of the quantization procedure is rather unproblematic, and the self-adjoint momentum shift operator, 
\begin{equation}
	\hat \pi_\varphi = -i\hbar \diby{}{\varphi}\,,
\end{equation}
conjugate to the point-wise multiplication operator, $\hat \varphi = \varphi$, can readily be constructed. This can be done because the $\varphi$-dependence of $\eta_{AB}$ is identical to that of the reference metric $\eta^0_{AB}$ on the conformal completion. Thus, as far as the $\varphi$-dependence of the wavefunction is concerned, the condition \eqref{eq:si gen} reduces to
\begin{equation}\label{eq:sq-int varphi}
	\Psi(\varphi \to \pm \infty) \to 0\,,
\end{equation}
which is the standard condition for L$^2(\mathbbm R, 1)$ functions. Moreover, the self-adjointness condition \eqref{eq:p s.a. ness} takes exactly the same form, and confirms the usual result that the translation operator on $\mathbbm R$ is self-adjoint.

For the directions orthogonal to the boosts, the situation is more subtle. We consider first a parametrization of this direction given by the volume $v$. For the momentum operator,
\begin{equation}
	\hat\pi_v \Psi \equiv \frac 1 {\sqrt v} \diby{}{v}\lf( \sqrt v \Psi \rt)\,,
\end{equation}
conjugate to the point-wise multiplication operator $\hat v = v$, the condition \eqref{eq:p s.a. ness} takes the form
\begin{align}
	\lf. v \Phi^\dag \Psi \rt|_{v = 0} & = 0 & \lf. v \Phi^\dag \Psi \rt|_{v \to \infty} & \to 0\,, 
\end{align}
because, $\sqrt{-\eta} = v$. This implies, in particular, that all wavefunctions, $\Psi$, satisfying the self-adjointness requirement can have a divergence of order up to (but not including) $v^{-1/2}$ at $v = 0$ because $\lf.\lf(\sqrt v |\Psi| \rt)\rt|_{v = 0} \to 0$. On the other hand, the condition for square integrability \eqref{eq:si gen} can be easily computed in this chart by noting that 
\begin{equation}
	\eta^0_{AB} = v^{-2} \eta_{AB}\,.
\end{equation}
Using this, we find
\begin{equation}\label{eq:sq_int cond}
	\lf.\lf(v |\Psi| \rt)\rt|_{v = 0} \to 0\,.
\end{equation}
This condition is \emph{weaker} than the self-adjointness requirement for the momentum operator conjugate to this coordinate because divergences of up to (but not including) order $v^{-1}$ are allowed. Thus, $\hat \pi_v$ is \emph{not} self-adjoint.

A well-defined momentum operator can be found by making use of the tortoise coordinate, $\mu$. In these coordinates, the self-adjointness requirement takes the form
\begin{equation}
	\lf. \lf( e^{\mu} |\Psi | \rt) \rt|_{\mu \to -\infty} \to 0\,,
\end{equation}
because $\sqrt{-\eta} = e^{2\mu}$. This is identical to the condition for square integrability. Thus, the momentum operator, $\hat\pi_\mu$, is essentially self-adjoint. This establishes the existence of a well-defined momentum operator conjugate to the tortoise operator, $\hat\mu = \mu$. An orthonormal basis for the spectrum of all momentum operators that fall into this class can be constructed using the reference metric via
\begin{equation}
	\psi_{k_A} = \frac 1 {2\pi \hbar} \lf(\frac{\eta_0}{\eta}\rt)^{1/4} e^{- \tfrac i \hbar q^Ak_A}\,.
\end{equation}
Note that the momentum of all the independent coordinates of this chart are included in this definition. One can easily verify that the functions above will span the space of functions that obey both the condition \eqref{eq:si gen} and the eigenvalue equation
\begin{equation}
	\hat p_A \psi_{k_A} = k_A \psi_{k_A}\,.
\end{equation}
For the tortoise coordinates, $(\mu, \varphi)$, for which the momentum operators are self-adjoint, these eigenfunctions take the explicit form
\begin{equation}\label{eq:mu-varphi eigenvalue eqn}
	\psi_{(k,r)} \equiv \frac 1 {2\pi \hbar} e^{-\mu + \tfrac i \hbar \lf( k \varphi +  r \mu \rt) }\,,
\end{equation}
where $(r,k)$ are the eigenvalues of $(\hat \pi_\mu, \hat \pi_\varphi)$ respectively. For the configuration operators that are represented by point-wise multiplication operators, self-adjointness is manifest and the spectrum is straightforwardly evaluated.

\subsection{Hamiltonian} 
\label{sec:hamiltonian}

Our next task is to construct self-adjoint representations of our Hamiltonian operator. A coordinate-invariant and symmetric representation of the Hamiltonian operator\footnote{As was emphasised in \S\ref{RelQ}, this Hamiltonian will be the appropriate one to use even when the cosmological constant is non-zero because $\Lambda$ can be interpreted as a separation constant obtained from solving our evolution equation similar to the total energy in a time-dependent Schr\"odinger equation.} corresponding to \eqref{eq:class H} can be written as:
\begin{equation}\label{eq:gen Ham}
	\hat H = - \frac {1} 2 \Box\,,
\end{equation}
where
\begin{equation}
	\Box = \frac{1}{\sqrt {-\eta} } \partial_{A} \lf( \eta^{AB} \sqrt{-\eta} \partial_{B} \rt)\,.
\end{equation}
is the standard Laplace--Berltrami operator on $\mathcal C$. This construction is similar to the procedure proposed in \cite{DeWitt:1957}. By virtue of being a real scalar function on $\mathcal C$, the Hamiltonian operator enjoys certain desirable properties not shared by the momentum operators of the previous section. Being scalar means that the self-adjointness condition will transform on $\mathcal C$ as a scalar in the same way as the square integrability condition. This further implies that different charts of $\mathcal C$ correspond to unitarily equivalent representations of the same operator. Moreover, being real means that the Hamiltonian is guaranteed to have self-adjoint extensions -- although these are not guaranteed to be unique. We will find below that the spectrum of the Hamiltonian operator splits into two qualitatively different branches: unbound states, which behave like bounce cosmologies and dominate early- and late-time solutions, and bound states, which are only important near the bounce and provide the potential for further phenomenological investigations for near-bounce physics.

The condition for self-adjointness of the Hamiltonian can be obtained by computing the boundary term in the integral below via two applications of integration by parts
\begin{multline}\label{eq:Ham bdy}
	\mean{\Phi, \hat H \Psi} = \mean{\hat H \Phi, \Psi}  - \frac 1 2 \oint_{\partial \mathcal C} \de l \sqrt{\chi}\, \eta^{AB} \\ \times n_{(A} \lf( \Phi^\dag \partial_{B)} \Psi - \Psi \partial_{B)} \Phi^\dag \rt)\,,
\end{multline}
where round brackets indicate symmetrization of indicies. Because this is a scalar condition, it is sufficient to work out the conditions for its vanishing in a global coordinate chart. The general conditions can be worked out in any chart via diffeomorphism. For a convenient chart, we will once again take advantage of the boost symmetry of the theory using the boost parameter, $\varphi$. It will turn out that the physically relevant variable $v$ will be most convenient for parametrizing the direction orthogonal to the boosts. In these variables, the metric is diagonal and the boundary term above splits into two pieces. The first piece is the boundary contribution in the limit where $|\varphi|$ grows unboundedly. In this limit, we have
\begin{equation}
	\lf.\lf(\Phi^\dag \partial_\varphi \Psi - \Psi \partial_\varphi \Phi^\dag \rt)\rt|_{\varphi \to \pm \infty} \to 0\,,
\end{equation}
which is automatically satisfied for all L$^2(\mathbbm R, 1)$ functions according to the condition \eqref{eq:sq-int varphi}. As this corresponds to the free-particle Hamiltonian on $\mathbbm R$, it is straightforward to see that the $\varphi$-part of the Hamiltonian is essentially self-adjoint.

The second contribution to the boundary term in \eqref{eq:Ham bdy} comes from the classically singular boundary $\partial \mathcal C$ at $v = 0$. This contribution takes the form
\begin{equation}\label{eq:s.a. ham}
	v \lf.\lf( \Phi^\dag \partial_v \Psi - \Psi \partial_v \Phi^\dag \rt)\rt|_{v = 0} = 0\,.
\end{equation}
We see that, provided the $v$-derivatives are regular at $v=0$, all functions satisfying the square integrability condition \eqref{eq:sq_int cond} will automatically satisfy this self-adjointness condition. Unlike the boundary contributions in the $\varphi \to \pm \infty$ limit, the solutions of \eqref{eq:s.a. ham} are only mutually orthogonal when they satisfy a unitary family of boundary conditions, and, thus, particular solutions are required to satisfy one member of these boundary conditions. A variety of techniques can be used to study these solutions and the nature of the boundary conditions in question. We will explore an avenue particularly catered to our problem in the sub-sections below. First, however, we will deal with the $\varphi$-dependence of the wavefunction.

The boost symmetry of $\hat H$ can be used to motivate a separation Ansatz for computing the square integrable eigenstates of $\hat H$ and its corresponding spectrum. If we postulate
\begin{equation}
	\Psi_{\Lambda}^\pm (v, \varphi) = \psi_{\Lambda, k}(v) \nu_{k}^\pm (k)
\end{equation}
as the linearly independent states satisfying the eigenvalue equation
\begin{equation}
	\hat H \Psi^\pm(v, \varphi) = \tilde \Lambda \tilde\Psi^\pm(v, \varphi)\,,
\end{equation}
then the $\varphi$-dependence of this equation can be found to satisfy
\begin{equation}\label{eq:wave}
	 \frac{\de^2}{\de \varphi^2} \nu_k^\pm = - \frac{k^2}{\hbar^2} \nu_k^\pm\,,
\end{equation}
while the $v$-dependence leads to
\begin{equation}\label{eq:bessel}
	v \frac{\de }{\de v}\lf( v \frac {\de}{\de v}  \psi_{\Lambda,k} \rt) + \lf( 2 \tilde \Lambda v^2 + \frac{k^2}{\hbar^2} \rt) \psi_{\Lambda,k} = 0\,.
\end{equation}
Equation \eqref{eq:wave} is the wave-equation equation whose solutions are the in- and out-going waves
\begin{equation}
	\nu_k^\pm(\varphi) = \frac {1}{\sqrt{2\pi\hbar}} e^{\pm \tfrac i \hbar k \varphi}\,.
\end{equation}
The square integrability condition \eqref{eq:sq-int varphi} is satisfied iff $k$ is real, and, therefore, the deficiency subspace has dimension zero, and the operator is essentially self-adjoint in line with our previous expectations.

Equation \eqref{eq:bessel} is Bessel's differential equation for purely imaginary orders, $ik$, given the restriction to real $k$. This equation has a regular singularity at $v= 0$, where we find physically interesting contributions to the boundary term \eqref{eq:s.a. ham}. Unlike the case of real or vanishing orders, none of the linearly independent solutions of \eqref{eq:bessel} for any eigenvalue is recessive at the origin. This means that there is no preferred choice of self-adjoint extension for $\hat H$. The character of the solutions to \eqref{eq:bessel} and the nature of the self-adjoint extensions differ significantly depending on the sign of the cosmological constant. We therefore treat each case separately below. The properties of the solutions used in our discussion, including the relevant asymptotic expansions, are described in detail in \cite{dunster1990bessel}, which we will reference in the following sub-sections. Alternatively, one can exploit the mathematical equivalence of our model with certain limiting regimes of the $1/r^2$ potential. Complimentary treatments of such potentials can be found, for example, in \cite{gopalakrishnan:2006,kunstatter:2009,barbour:2013}.

\subsubsection{`Bound States' ($\Lambda < 0$)} 
\label{sub:bound states}

When the cosmological constant is negative, the linearly independent solutions to \eqref{eq:bessel} are the modified Bessel functions of the first, $I_{ik/\hbar}(\sqrt{2\tilde \Lambda} v)$, and second kind, $K_{ik/\hbar}(\sqrt{2\tilde \Lambda} v)$. The asymptotic form of these functions for large $x = \sqrt{2\tilde \Lambda} v$ is given by
\begin{align}
	I_{ik/\hbar}(x) &\sim \frac {e^{x}}{\sqrt{x}}(1 + \mathcal O(1/x)) \\
	K_{ik/\hbar}(x) &\sim \frac {e^{-x}}{\sqrt{x}}(1 + \mathcal O(1/x)) \,.
\end{align}
The exponential growth of $I_{ik/\hbar}(x)$ is not compatible with the square integrability requirement in the $v\to \infty$ limit. Thus, the only potentially square integrable solution is $K_{ik/\hbar}(x)$, which decays exponentially in $v$ and, thus, represents a state `bound' to the singular region near $v = 0$. We will henceforth refer to such states as `bound states'. To examine the square integrability and self-adjointness of $K_{ik/\hbar}(x)$, we need to examine its behaviour near the regular singularity at $v = 0$. This is given by
\begin{equation}
	K_{ik/\hbar}\lf(\sqrt{2\tilde \Lambda}v\rt) \sim \sin\lf( \tfrac k \hbar \log v + \tfrac k {2\hbar} \log \lf( \tfrac {\tilde \Lambda} 2 \rt) - \theta_k \rt)\,,
\end{equation}
where $\theta_k =  \text{Arg}(\Gamma(ik/\hbar + 1))$ and $\Gamma$ is the Gamma-function. The limit is not well-defined at $v = 0$, but the result is bounded and, therefore, square integrable in accordance with \eqref{eq:sq_int cond}. Moreover, it is possible to assess whether these solutions are in the self-adjoint domain of $\hat H$ by considering their behaviour in \eqref{eq:s.a. ham} for different values of the cosmological constant -- say $\Lambda_a$ and  $\Lambda_b$. After inserting the asymptotic expansion above into \eqref{eq:s.a. ham} for $\Phi = K_{ik/\hbar}(\sqrt{2\tilde \Lambda_a} v)$ and $\Psi = K_{ik/\hbar}(\sqrt{2\tilde \Lambda_b} v)$, we obtain
\begin{equation}
	\sin \lf[ \tfrac k {2\hbar} \log\lf( \Lambda_a/\Lambda_b \rt) \rt] = 0\,.
\end{equation}
The zeros of this function occur, for integer $n$, when
\begin{equation}
	\frac {\Lambda_a}{\Lambda_b} = e^{2n\pi\hbar/k}\,.
\end{equation}
This implies that the bound eigenstates\footnote{The normalization below is computed by requiring that the states be orthonormal and by making use of the integral: $ \int_0^\infty \de x\, x K_{ik}(u_a x)K_{ik}(u_b x) = \frac{\pi k}{2u_a u_b} \delta_{ab} $.}
\begin{equation}
	\psi_{\Lambda,k}^\text{bound} =  \sqrt{\frac {4\hbar |\tilde \Lambda| \sinh \lf(\pi k/\hbar\rt)} {\pi k} }\, K_{ik/\hbar}(\sqrt{2 \tilde \Lambda } v)
\end{equation}
belong to the self-adjoint domain of $\hat H$ provided the spectrum is restricted to discrete values of $\Lambda$ that differ by factors of $e^{2n\pi\hbar/k}$. The particular set of discrete values to be used represents a particular choice of self-adjoint extension of the domain of $\hat H$. This choice can be parametrized by a choice of reference scale $\Lambda_\text{ref}$, which is under-determined up to a log-periodicity given by $\log \Lambda_\text{ref} \to \log \Lambda_\text{ref} + 2\pi\hbar/k$. Within a particular choice of self-adjoint extension, it is straightforward to verify that, given the normalization defined above, the orthogonality relations
\begin{equation}
	\int_0^\infty \de v v\, \psi^\dag_{\Lambda_a,k}, \psi_{\Lambda_b,k} = \delta_{ab}\,
\end{equation}
hold in accordance with the observation that Bessel's equation, \eqref{eq:bessel}, is of Sturm--Liouville form. These can be used to construct general states. Combining these eigenfunctions with the arbitrary linear combinations of the $k$-space eigenfunctions, we can construct a general $\Lambda$-eigenstate in terms of
\begin{widetext}
\begin{equation}\label{eq:ads eigenfunctions}
	\Psi^\text{bound}_\Lambda(v, \varphi) = \sqrt{\frac {2} {\pi^2 \hbar}} \int_{0}^\infty \de k\, ( A(k) \cos ( \tfrac {k\varphi} \hbar )  + B(k) \sin ( \tfrac {k\varphi} \hbar ) ) \sqrt{ \frac { |\tilde \Lambda| \sinh ( \pi k/\hbar )}k }  K_{i k/\hbar} ( \sqrt{2 |\tilde \Lambda |} v )\,,
\end{equation}
\end{widetext}where the normalization $\int_{0}^{\infty} \tfrac {\de k}\hbar |A(k)|^2 = \int_{0}^{\infty} \tfrac {\de k}\hbar |B(k)|^2 = 1$ guarantees
\begin{equation}
	\mean{\Psi^\text{bound}_{\Lambda_a}, \Psi^\text{bound}_{\Lambda_b} } = \delta_{ab}\,.
\end{equation}
The oddness of $K_{ik/\hbar}$ in terms of $k$ projects out the even part of the $k$-space solution and allows \eqref{eq:ads eigenfunctions} to be written as an integral in $k$ from $0$ to $\infty$.

This spectrum is not bounded from below and has an accumulation point at $\Lambda = 0$. Nevertheless, states of this kind can be constructed as effective field theories in atomic 3-body systems, and the resulting phenomenon is known as the \emph{Efimov effect} \cite{efimov:1970,gopalakrishnan:2006,ferlaino:2010}. The literature on the Efimov effect is vast. However, this is the first time, to our knowledge, that a relation has been pointed out between this effect and the potential physics of the early Universe. The implications for this will be discussed in Section \S\ref{analogue}. 

\subsubsection{`Unbound States' ($\Lambda > 0$)} 
\label{sub:unbound states}

When the cosmological constant is positive, the linearly independent solutions to \eqref{eq:bessel} are the Bessel functions of the first, $J_{ik/\hbar}(\sqrt{2\tilde \Lambda}v)$, and second, $Y_{ik/\hbar}(\sqrt{2\tilde \Lambda}v)$, kind. In the case of purely imaginary order, these functions need to be rescaled so that they are numerically well-behaved and suitably normalized functions of $k$. A choice of normalization that is both real and well-behaved is given by (see \cite{dunster1990bessel})
\begin{align}
	\mathcal F_{k}(x) &\equiv \frac 1 2 \text{sech} \lf( \frac {\pi k}{2\hbar} \rt) \lf[ \mathcal J_{i k/\hbar}(x) + \mathcal J_{-i k/\hbar}(x) \rt]\\
	\mathcal G_{k}(x) &\equiv \frac 1 {2i} \text{cosech} \lf( \frac {\pi k}{2\hbar} \rt) \lf[ \mathcal J_{i k/\hbar}(x) - \mathcal J_{-i k/\hbar}(x) \rt]\,.
\end{align}
For sufficiently large $v$, these functions behave like oscillating functions with a $1/\sqrt{v}$ decay envelope:
\begin{align}
	\mathcal F_{k}(x) &\approx \lf( \frac 2 {\pi x} \rt)^{1/2} \lf[ \cos\lf( x - \pi/4 \rt) + \mathcal O(1/x) \rt] \label{eq:F decay}\\
	\mathcal G_{k}(x) &\approx \lf( \frac 2 {\pi x} \rt)^{1/2} \lf[ \sin\lf( x - \pi/4 \rt) + \mathcal O(1/x) \rt]\,.\label{eq:G decay}
\end{align}
Note the $k$-independence of the leading order behaviour, which has resulted from the choice of normalization of these functions. These eigenfunctions satisfy the square integrability and self-adjointness requirements for $v\to \infty$. This is not automatically the case, however, on the other branch of the boundary at the regular singularity at $v = 0$.

In this case, the Bessel functions behave as
\begin{multline}
	\mathcal F_{k}\lf(\sqrt{2\tilde \Lambda} v\rt) \approx \lf( \frac{2\hbar \tanh (\pi k/2\hbar)}{\pi k} \rt)^{1/2} \\ \times \lf( \cos\lf[ \tfrac k \hbar \log v + \tfrac k {2\hbar} \log \lf( \tfrac {\tilde \Lambda} 2 \rt) - \theta_k \rt]  + \mathcal O(v^2) \rt) 
\end{multline}
and	
\begin{multline}
	\mathcal G_{k}\lf(\sqrt{2\tilde \Lambda} v\rt) \approx \lf( \frac{2\hbar \coth (\pi k/2\hbar)}{\pi k} \rt)^{1/2} \\ \times \lf( \sin\lf[ \tfrac k \hbar \log v + \tfrac k {2\hbar} \log \lf( \tfrac {\tilde \Lambda} 2 \rt) - \theta_k \rt]  + \mathcal O(v^2) \rt)\,.
\end{multline}
As in the case of the bound states, the $v \to 0$ limit is not well-defined. Nevertheless, the functions are bounded and, therefore, the square integrability condition is satisfied. To solve the self-adjointness condition, we first note that real functions automatically obey \eqref{eq:s.a. ham} when $\Phi = \Psi$. Since we have already constructed two real, linearly independent solutions we can construct a new solution, for a single value of $\Lambda$, by adding them in a unitary combination such as $\psi_\theta = \cos \theta \psi_1 + \sin \theta \psi_2$. To get the full solution, we need to find states such as these that independently solve the boundary conditions for different values of $\Lambda$. To deal with this, it is sufficient to note that the logarithmic dependence of the arguments of the trigonometric functions above makes it possible to subtract the $\Lambda$-dependence of these functions near the origin by inserting the correct relative phase between the two linearly independent solutions. To this end, it is convenient to take linear combinations of the $\mathcal F_k$'s and the $\mathcal G_k$'s to form $\mathcal J_{ik/\hbar}$'s and note that the combination $(\Lambda/\Lambda_\text{ref})^{ik/2\hbar}$ is sufficient to cancel the $\Lambda$-dependence of the $\mathcal J_{ik/\hbar}$'s near the $v = 0$ boundary for any value of the reference scale $\Lambda_\text{ref}$. The most general solution is to take unitary linear combinations of real and imaginary parts. A convenient way to keep track of this freedom is expressed through the use of the $k$-dependent reference scale $\Lambda_\text{ref}(k)$. Different choices of $\Lambda_\text{ref}(k)$ can be made to correspond to different phases between real and imaginary pieces and can, therefore, be used to parameterize the different possible choices of self-adjoint extensions. The most general solution, therefore, is given by:\footnote{ Note that another useful relation for writing the normalization is: $  \sqrt{ \cosh \lf( \frac{\pi k}\hbar \rt)  + \cos \lf( \frac k \hbar \log \lf[\tfrac \Lambda {\Lambda_\text{ref}}\rt] \rt) } = \sqrt 2 \lf| \cosh \lf( \tfrac {\pi k}{2\hbar} + i \tfrac k {2\hbar} \log\lf[ \tfrac {\Lambda}{\Lambda_\text{ref}} \rt] \rt)  \rt| $. }
\begin{equation}
	\psi^{\Lambda_\text{ref}}_{\Lambda,k} 
	= \frac{ \text{\cal{Re}} \lf[  \lf( \frac {\Lambda} {\Lambda_\text{ref}(k)} \rt)^{-ik/2\hbar} \mathcal J_{ik/\hbar}(\sqrt{2 \tilde \Lambda} v) \rt]}{ \lf| \cosh \lf( \tfrac {\pi k}{2\hbar} + i \tfrac k {2\hbar} \log\lf[ \tfrac {\Lambda}{\Lambda_\text{ref}(k)} \rt] \rt)  \rt|  }\,.
\end{equation}
The normalization is chosen so that the eigenstates satisfy
\begin{equation}
  \int_0^\infty \de v v \, \lf(\psi^{\theta}_{\Lambda_a, k}\rt)^\dag \psi^{\theta}_{\Lambda_b, k} = \delta ( \tilde \Lambda_a - \tilde \Lambda_b )\,,
\end{equation}
which can be verified by inserting the large $v$ behaviour of the $\mathcal J$'s. Note that simple choices for $\Lambda_\text{ref}$ could be the mean of $\hat \Lambda$ or a constant phase, $\alpha$, such that $\Lambda_\text{ref}(k) = e^{-\alpha k/2\hbar}$. Note that, the imaginary piece can be obtained through a suitable redefinition of $\alpha$.

With this choice of wavefunction, the near boundary eigenstate takes the form
\begin{equation}
	\psi_{\Lambda, k}^\theta(v) \propto \cos\lf[ \tfrac k \hbar \log v + \tfrac k {2\hbar} \log \lf( \frac {\tilde \Lambda} 2 \rt) - \theta_k - \theta \rt]\,,
\end{equation}
where
\begin{equation}
	\theta = \frac k {2\hbar} \log\lf( \frac \Lambda {\Lambda_\text{ref}} \rt)\,.
\end{equation}
This satisfies the boundary condition \eqref{eq:s.a. ham} by construction. The $U(1)$ self-adjoint extension parameter $\theta$ gives a more precise representation of the physics of the dimensionful reference scale $\Lambda_\text{ref}$. In particular, we note that the periodicity $\theta \to \theta + n\pi$ (for integer $n$) implies $\psi_{\Lambda, k}^{\Lambda_\text{ref}}(v) \to - \psi_{\Lambda, k}^{\Lambda_\text{ref}}(v)$, which is not a linearly independent function. This means that the reference scale is only defined up to
\begin{equation}
 	\Lambda_\text{ref} \to e^{2n\pi\hbar/k} \Lambda_\text{ref}\,,
 \end{equation} 
as was the the case for $\Lambda< 0$.

The physical interpretation of $\theta$ can be seen by evaluating the eigenfunction $\psi^{\Lambda_\text{ref}}_{\Lambda,k}$ in the large $v$ limit. In this limit, the asymptotic forms \eqref{eq:F decay} and \eqref{eq:G decay} can be used to show that
\begin{multline}
	\psi^{\Lambda_\text{ref}}_{\Lambda,k} \propto \cos\lf( \sqrt{2\tilde \Lambda}v - \tfrac \pi 4 - \arctan \lf( \tanh \lf({\tfrac {\pi k}{2\hbar}}\rt) \tan \theta \rt) \rt) \\ + \mathcal O\lf(1/(\sqrt{2\tilde \Lambda}v)\rt)\,.
\end{multline}
Since the cosine function is a linear combination of in- and out-going modes with opposite phases, this implies a phase shift of
\begin{equation}\label{eq:phase shift}
	\Delta \equiv \frac \pi 2 + 2 \arctan \lf[ \tanh \lf({\tfrac {\pi k}{2\hbar}}\rt) \tan \lf( \tfrac k {2\hbar} \log\lf( \tfrac \Lambda {\Lambda_\text{ref}} \rt) \rt) \rt]
\end{equation}
between asymptotic planes waves as they scatter in and out of the bounce regime.  Note that the log periodicity of $\Lambda_\text{ref}$ implies that the $\Lambda_\text{ref} \to 0$ and $\Lambda_\text{ref} \to \infty$ limits are not well-defined. This means that $\Delta$, according to \eqref{eq:phase shift}, cannot take the value zero and, therefore, that there is no preferred choice of self-adjoint extension. The phase shift $\Delta$ also defines, via the moments of the wavefunction, a preferred notation of temporal units for the system via a characteristic time for deep scattering. These considerations will be crucial for interpreting the physics of the cosmological solutions studied in the companion paper \cite{Gryb:2017b}.

If $\Lambda_\text{ref}(k)$ where independent of $k$ (or even in $k$), then the unbound wavefunctions would be even in $k$ in contrast to the bound wavefunctions that were odd in $k$. The most general case, however, for the bound $\Lambda$-eigenstates involves arbitrary superpositions of positive and negative values of $k$:
\begin{widetext}
\begin{equation}\label{eq:ds eigenstates}
	\Psi^{\text{unbound}}_{\Lambda, \Lambda_\text{ref}} = \int_{-\infty}^\infty \de k\, \frac{ C(k) \cos \lf( \tfrac {k\varphi}{\hbar} \rt) + D(k) \sin \lf( \tfrac {k\varphi}{\hbar} \rt)}{\sqrt{2 \pi}\hbar\lf| \cosh \lf( \tfrac {\pi k}{2\hbar} + i \tfrac k {2\hbar} \log\lf[ \tfrac {\Lambda}{\Lambda_\text{ref}} \rt] \rt)  \rt|} \text{\cal{Re}} \lf[  \lf( \frac {\Lambda} {\Lambda_\text{ref}} \rt)^{-ik/2\hbar} \mathcal J_{ik/\hbar}(\sqrt{2 \tilde \Lambda} v) \rt]\,,
\end{equation}\end{widetext}
where the normalization $\int_{-\infty}^\infty \frac{\de k}\hbar\, |C|^2 = \int_{-\infty}^\infty \frac{\de k}\hbar\, |D|^2 = 1$ guarantees
\begin{equation}
	\mean{ \Psi^{\text{unbound}}_{\Lambda_a, \Lambda_\text{ref}}, \Psi^{\text{unbound}}_{\Lambda_b, \Lambda_\text{ref}} } = \delta ( \tilde\Lambda_a - \tilde\Lambda_b )\,.
\end{equation}

As a final comment on the interpretation of $\Lambda_\text{ref}$, we note that the vanishing of the boundary term in \eqref{eq:Ham bdy} is a conformally invariant equation because the d'Alembertian is conformally covariant in 2d. It should, thus, be no surprise that the solutions to this equation take the form of the eigenfunctions of $\Box$ in the conformal completion $(\mathcal C_0, g_0)$. One can, therefore, understand our procedure as a way of anchoring the Bessel functions using the known solutions in on the conformally completed manifold. The introduction of the dimensionful parameter $\Lambda_\text{ref}$ can thus be interpreted as a necessary ingredient for removing the dependence of the bulk eigenfunctions on the dimensionful parameters of the theory so that the bulk theory can be compatible with the conformal boundary condition.


\subsubsection{Critical Case ($\Lambda = 0$)} 
\label{sub:subsection_name}

The case where $\Lambda$ is precisely vanishing is a critical point\footnote{Discussions of the link to critical phenomena in the context of $1/r^2$ potentials can be found, for example, in \cite{gopalakrishnan:2006,kunstatter:2009,barbour:2013}.} of our model. The eigenvalue equation for the Hamiltonian becomes conformally invariant since the Laplacian is conformally covariant in 2d. The flat reference metric $\eta_0$ on the conformal completion of $\mathcal C$ can then be equivalently used to construct the quantum theory. In terms of the conformal completion, the Hamiltonian becomes the flat metric on $\mathbbm R^{1,1}$. The degenerate spectrum, in terms of the tortoise coordinates $\mu$ and $\varphi$, consists of in- and out-going plane waves on the $\mu\varphi$-plane satisfying a dispersion relation $r^2 = k^2$ (in terms of the eigenvalues defined in \eqref{eq:mu-varphi eigenvalue eqn}). The domain of the Hamiltonian for any particular choice of $\Lambda_\text{ref}$, which is essentially self-adjoint in this representation, is then given by standard $L(\mathbbm R^2, 1)$ functions.

Although the Hamiltonian is well-defined and anomaly-free, the quantization in this regime is disconnected from the quantum solutions given above for non-zero $\Lambda$. This is because the solutions of the former cannot be obtained as a continuous limit of the latter since the Bessel functions are not defined in this limit. Hence, the $\Lambda = 0$ case should be regarded as a separate quantization confined to the single eigenvalue $\Lambda = 0$. This solution, however, leads to a timeless wavefunction because the phase $e^{i\Lambda t/\hbar} = 1$.\footnote{The difference between $\Lambda>0$ and $\Lambda=0$ solutions is analogous to the difference between massive $m>0$ and massless $m=0$ particles. The latter solutions (which must be light-like) are not obtained from the former (which must be time-like) in a smooth limit.} We, therefore, do not regard this as a physically interesting cosmological solution in the relational quantization paradigm.

 
\subsection{Bouncing Unitary Cosmology}\label{SE for U}

Application of the relational quantization procedure of \S\ref{RelQ} leads directly to a Schr\"odinger-type equation of the form:
\begin{equation}\label{eq:main ev}
	\hat H \Psi = i\hbar \diby \Psi t\,,
\end{equation}
where we use the Hamiltonian operator \eqref{eq:gen Ham}. Since the Hamiltonian is self-adjoint, the time evolution is guaranteed to be unitary by  Stone's theorem \cite[p.264]{reed:1980}.

It is important to note that in this formalism, $t$ is treated as an arbitrary label that orders successive instants. It is not a physical observable. Rather, all observations can be transcribed purely in terms of the relative changes in correlations between operators. In this case, a complete set of self-adjoint operators that generalize the full classical phase space are the operators $ \hat q^A,\hat p_A$. As outlined in \S\ref{sec:observables}, a self-adjoint representation of these operators can be given explicitly in terms of $\hat \mu, \hat\varphi, \hat \pi_\mu$, and $\hat \pi_\varphi$ as in \S\ref{sec:observables}. Although, $t$ is needed as a bookkeeping device to label how the observables change relative to each other, the actual value of $t$ is never required to be directly or indirectly measured.

Having constructed an explicit representation for our Hilbert space and computed the self-adjoint spectrum for our Hamiltonian, it remains to explicitly write the most general solution to \eqref{eq:main ev}. A separation Ansatz
\begin{equation}
	\Psi(q^A, t) = \int_{-\infty}^\infty \de \tilde\Lambda\, E(\tilde\Lambda) e^{-\tfrac i \hbar \tilde\Lambda t} \Psi_\Lambda
\end{equation}
reduces \eqref{eq:main ev} to the eigenvalue equation:
\begin{equation}
	\hat H \Psi_\Lambda = \tilde \Lambda \Psi_\Lambda\,,
\end{equation}
which we have already solved in detail in \S\ref{sec:hamiltonian}. The general solution can constitute a superposition of negative bound $\Lambda$-eigenstates and positive unbound $\Lambda$-eigenstates. However, because the bound states decay exponentially away from the bounce region at $v=0$, they necessarily represent re-collapsing cosmologies and, therefore, appear to be in contradiction with observations. Contrastingly, the unbound states have an oscillating norm at large $v$ and therefore have the potential to reproduce a late-time expanding FLRW universe with positive cosmological constant in line with current observations. Moreover, while the unbound states admit continuous values of $\Lambda$ and have well-defined semi-classical limit for all choices of the parameter space, the bound states are discrete in $\Lambda$ and only admit a semi-classical limit when the parameter space is tuned in such a way that the wavefunction can have support in regions of comparatively large $v$ and $\varphi$. Nevertheless, the bound wavefunction could have important contributions in the deep quantum regime when the bound and unbound wavefunctions have significant overlap. In the companion paper \cite{Gryb:2017b}, we explore the potential implications of this potential period of overlap for cosmological observations.

In terms of these considerations, the general solution can then be taken to be
\begin{multline}\label{eq:gen soln}
	\Psi(q^A, t) = \frac 1 {\sqrt 2} \int_{0}^\infty \de \tilde\Lambda\, \lf[ e^{\tfrac {i} \hbar \tilde\Lambda t} E^-(\tilde\Lambda) \Psi^\text{bound}_{-\Lambda} \rt. \\ \lf. + e^{-\tfrac {i} \hbar \tilde\Lambda t} E^+(\tilde \Lambda) \Psi^\text{unbound}_{\Lambda,\Lambda_\text{ref}}  \rt]\,,
\end{multline}
where the bound and unbound eigenfunctions are respectively given by \eqref{eq:ads eigenfunctions} and \eqref{eq:ds eigenstates} and we take the functions $E^{\pm}$ to be normalized such that $\int_0^\infty \de \tilde \Lambda |E^\pm|^2 = 1$. Given our discussion in \S\ref{sub:bound states}, the bound eigenstates admit only a discrete spectrum specified by the choice of self-adjoint extension parameter, which can be parameterized by $\Lambda_\text{ref}$. We, therefore, find that the $\Lambda$-dependence of $E^-(\tilde\Lambda)$ can only take discrete values, and the above integral reduces to the sum
\begin{multline}
	\Psi(q^A, t) = \frac 1 {\sqrt 2} \lf[  \sum_{n= -\infty}^\infty e^{i \tilde \Lambda_n t/\hbar} E_n \Psi^\text{bound}_{-\Lambda_n}\rt. \\ + \lf. \int_{0}^\infty \de \Lambda\, e^{ -i \tilde\Lambda t/\hbar } E^+(\tilde \Lambda) \Psi^\text{unbound}_{\Lambda,\Lambda_\text{ref}}  \rt]\,,
\end{multline}
where $n \in \mathbbm Z$, $\sum_n |E_n|^2 = 1$ and
\begin{equation}
	\Lambda_n = \Lambda_\text{ref}^{2n\pi\hbar/k}\,.
\end{equation}

\subsection{Efimov Analogue Cosmology}\label{analogue}

As was noted above, fascinating connections exists between the physics of our cosmological model and that of atomic 3-body systems. In particular, the bound eigenstates \eqref{eq:ads eigenfunctions} have a mathematical form that mirrors that of  certain bound atomic trimer states, the physics of which is described by an effective $1/r^2$ potential. These states are known in the literature as \emph{Efimov states} \cite{efimov:1970,gopalakrishnan:2006,ferlaino:2010}. Given the experimental realisation and manipulation of Efimov states using ultra-cold atoms \cite{kraemer:2006}, our proposal thus presents the opportunity for a novel and experimentally viable platform for analogue quantum cosmology. In general terms, analogue approaches to gravity \cite{barcelo:2005} are based upon a formal correspondence between the equations describing particular condensed matter systems, such as Bose-Einstein condensates, and the semi-classical description of gravitational systems such as black holes \cite{unruh:1981,garay:2000} and FLRW \cite{barcelo2003analogue} or inflationary cosmologies \cite{fischer2004quantum,uhlmann2005aspects,cha:2017}.\footnote{See \cite{cha:2017} for a more complete list of references on analogue cosmology.}. There has been exciting recent experimental progress in using analogue models to simulate aspects of Hawking radiation in the lab \cite{weinfurtner:2013,steinhauer:2014,steinhauer:2016}. Analogue models for bouncing cosmologies have been suggested before \cite{oriti:2017}. However, our proposed \textit{Efimov analogue cosmology} is entirely novel. The particular importance of the analogy for the present work is the interpretation of the self-adjoint extension parameters.  

Consider first the atomic system in the context of a scattering experiment where a plane wave is scattered off an Efimov state. Here the self-adjoint extension parameter is fixed by the fundamental physics of the bound state. That is, the true potential that governs the trimer atomic system in the regime when the effective $1/r^2$ Efimov model breaks down. Applying the self-adjoint extension procedure to the relevant Hamiltonian operator introduces a dimensionful parameter into the effective physics of the trimer sate. The value of this parameter can then be measured by its effect on the plane wave during the scattering. In particular, the extension parameter fixes the magnitude of a phase shift in the plane wave during the scattering process. Thus we see that in the atomic Efimov effect the extension parameter is a genuine physical magnitude that: i) has its origin in the fundamental 3-body atomic physics; and ii) can be measured via the phase shift experienced by an scattered plane wave. Given this, it is natural to consider an analogue version of this `experiment' for the Efimov cosmology.  That is, we can consider a `scattering' of the unbounded states off the bound states in the deep quantum early universe regime. In this sense appeal to Efimov analogue cosmology provides a very natural strategy for the physical interpretation of the self-adjoint extension parameter, $\theta$. As noted above, $\theta$ is more conveniently parametrised in terms of a reference value of the cosmological constant $\Lambda_\text{ref}$. In \eqref{eq:phase shift} we explicitly expressed the phase shift, $\Delta$, by considering the difference between the asymptotic forms of the constant $\Lambda$-eigenstates when $v$ is large, and where the amplitudes of these eigenstates behave like cosines. In keeping with the atomic analogy, this phase shift can now be interpreted as a giving a particular scattering length that parametrises the size of the region in which the model is no longer accurate. In the companion paper we provide an explicit analysis of the unbound states with the self-adjoint extension parameter constrained by the semi-classical value of the cosmological constant \cite{Gryb:2017b}. 

\section{Singularity Resolution by Quantum Evolution} \label{sec:sing res}

What does it mean for a classical cosmological singularity to be `resolved' in the quantum theory? One influential suggestion is that non-singular quantum behaviour is indicated by the boundedness of the inverse 3-volume operator \cite[p.296]{Rovelli:2004}, which indicates a potential degeneracy in the spacetime metric. Such a kinematic criterion is prima facie well motivated on the basis of the classical relation between curvature divergence and the existence of the inverse metric components. However, the boundedness of kinematical operators, such as the inverse of the 3-volume, is not alone sufficient to establish non-singular behaviour \cite{bojowald:2006,bojowald:2007}. Rather, one needs to additionally consider dynamics in order to relate bounded operators to non-divergent curvature.\footnote{The need to consider dynamics is made abundantly clear by the case of anisotropic cosmological models where non-singular evolution exists despite the relevant curvatures being not necessarily bounded \cite{bojowald:2007}.} 

An alternative dynamical criterion is in terms of Bojowald's notion of quantum hyperbolicity \cite{bojowald:2007}: a non-singular quantum spacetime obtains when a state can be uniquely extended across or around all submanifolds of classically singular configurations.\footnote{Bojowald's criterion is in the same spirit as the dynamical singularly resolution proposals outlined in \cite{husain:2004,Kiefer:2007}} Although well motivated for the analysis of quantum singularities in Wheeler--DeWitt theories with internal times, quantum hyperbolicity is not particularly relevant when applied to the analysis of singularities in the context of quantum cosmological models with unitary Schr\"odinger-type evolution. In the Wheeler--DeWitt case, the physical Hilbert space on which evolution takes place is represented in terms of operators acting on functions whose domains are ever changing submanifolds, parametrized by internal time, of the configuration space. Here, the dynamics can force the system directly through a classically singular submanifold, and this can lead to potential problems. In contrast, in the unitary case, the domain of the relevant functions remains fixed to the entirety of the original configuration space. Because the classically singular region is a set of measure zero on this space, a representation for bounded operators can always be found using the prescriptions developed in \S\ref{sec:observables}.

In our view, the basic condition for non-singular behaviour is that the expectation value of an observable operator, as evaluated on all possible state in Hilbert space, always remains finite. Given a classical theory in which some classical observable diverges it is necessary for singularity resolution that the expectation value of the corresponding quantum operator is always finite. This finite expectation value criterion is applicable both to Schr\"{o}dinger-type models, where the expectation values are considered at all values of the external time parameter, and to Wheeler--DeWitt-type models, where the expectation values are projected onto a physical Hilbert space and are considered at all values of some internal time. Finiteness of expectation values can understood as a special case of a more general requirement for the existence of well defined effective equations of motion as given by quantum corrections to Hamilton's equations for the expectation values \cite{bojowald:2006b,brizuela:2014}.\footnote{These corrections can be explicitly characterised in terms of the \textit{quantum moments} associated with the system. For analysis of Wheeler-DeWitt type quantum cosmology and Loop Quantum cosmology within the framework of effective theory based upon moment expansion see \cite{bojowald:2007b,bojowald:2011b,bojowald2012c}. We leave application of this framework to  Schr\"{o}dinger-type model for future work.}  Thus, in requiring finiteness of expectation values we are not implicitly relying upon extending the correspondence principle into the quantum bounce regime, but rather simply insisting that a physically reasonable quantum theory can be defined at all times. 


That our model satisfies the finite-expectation-values condition for singularity avoidance is straightforward to demonstrate. Given that the Hamiltonian is self-adjoint and bounded\footnote{It is bounded by virtue of restricting to square integrable function on its domain and remains bounded because it commutes with itself.} and we have a unitary Schr\"{o}dinger equation, the generalised Ehrenfest's theorem applies:
\begin{equation}\label{eq:Ehrenfest}
\frac{\partial}{\partial t} \Big{<}\hat{A}(t)\Big{>} = \frac{1}{i\hbar} \Big{<}[\hat{A}(t),\hat{H}]\Big{>}+ \Big{<}  \frac{\partial \hat{A}(t)}{\partial t}\Big{>}\,.
\end{equation} 
Provided the quantum observable algebra of bounded operators is well-defined, the commutator on the RHS is also bounded and the evolution of all observable expectation values will be well-behaved. Thus, for unitary quantum cosmology, the condition for singularity avoidance ultimately amounts to the usual requirement for a well-defined quantum theory. 


We can now explicitly contrast the fate of the big bang singularity in mini-superspace models under Wheeler--DeWitt-type and Schr\"{o}dinger-type quantizations. For the Wheeler--DeWitt case, the relevant expectation values can be been computed explicitly under the assumptions of a massless scalar field, zero spatial curvature, and zero cosmological constant \cite{ashtekar:2008}. Given the choice of $\phi$ as the internal clock, one can construct a one parameter family of self-adjoint Dirac observables $\hat{V}|_{\phi_{0}}$ each of which have the physical interpretation of the (spatial) volume operator at `time' $\phi=\phi_0$. In \cite[IV]{ashtekar:2008}  it is shown that for left moving states in the domain of the volume operator,  the expectation value $\big{<}\hat{V}|_{\phi_{0}} \big{>}\rightarrow\infty$ as $\phi\rightarrow\infty$ and $\big{<}\hat{V}|_{\phi_{0}} \big{>}\rightarrow0$ as $\phi\rightarrow-\infty$. The situation is reversed for the right moving sector. Thus, for a dense subset of physical states, Wheeler--DeWitt quantization fails our necessary condition to resolve the cosmic singularity. This explicitly illustrates what can go wrong in the WDW formalism according to our general discussion above: the deparametrization with respect to an internal clock, which must take explicit values, can force the dynamics of the system directly into the classically singular region where things can easily go wrong quantum mechanically. What is lacking in the Wheeler--DeWitt approach is an evolution with respect to a self-adjoint (or even symmetric) Hamiltonian. This fact bars the use of the Ehrenfest theorem, which guarantees that all operators remain bounded in the unitary approach.

\section{Conclusions}

The core result of this paper is the construction of a bouncing unitary cosmology from a classically singular mini-superspace model. We have shown that solutions to our universal Schr\"{o}dinger equation are generically singularity free and a set of well-defined quantum observables can be constructed with the corresponding expectation values guaranteed to be finite. Generic features of these solutions include: i) superpositions of values of the cosmological constant; ii) universal effective physics due to non-trivial self-adjoint extensions of the Hamiltonian; and iii) bound `Efimov universe' states for negative cosmological constant. In a companion paper, we will explore key features of specific solutions. We will find that solutions can be constructed which contain: a cosmic bounce due to quantum gravitational effects, a well-defined FLRW limit for $\Lambda > 0$ far from the bounce, and a semi-classical turnaround point in the dynamics of $\hat\varphi$ which resembles an effective inflationary epoch. This suggests several novel fronts for further investigations.

In more general terms, from our perspective, the quantization of singular cosmological models under the standard Wheeler-DeWitt approach is doubly problematic. First, there is the lack of unitary time evolution and second there is the persistence of pathological features, as indicated by non-finite expectation values for classically singular quantities. The alternative approach outlined is this paper is guaranteed to resolve both issues for any model to which it can be consistently applied. By requiring unitary Schr\"{o}dinger evolution of the universal wavefunction, as generated by a self-adjoint Hamiltonian operator, we provide a generic mechanism for singularity avoidance in quantum cosmology. Moreover, the interpretation of the self-adjoint extension parameter suggested by the analogue atomic systems is in terms of an effective low-energy parameter fixed by unknown high-energy physics. This could allow for a comparison between our model and other bounce approaches stemming from proposals for UV-completions of quantum gravity such as Loop Quantum Cosmology \cite{Bojowald:2008} or non-commutative geometry \cite{arraut:2010}.

This said, there are important limitations to our approach: in particular, the relational quantization procedure appealed to here is in its present form only applicable to models with a single Hamiltonian constraint, and is thus not readily extendable to models with spatial inhomogeneities. The prospects for extension of our approach to more realistic cosmological models thus depends upon   approaches to gravity in which the constraint structure of the Dirac's hyper-surface deformation algebroid is exchanged for a simpler structure featuring a single Hamiltonian constraint. An example of such an approach is given by the Shape Dynamics formulation of gravity \cite{barbour_el_al:physical_dof, Barbour:new_cspv, gryb:shape_dyn, Gomes:linking_paper}. As noted in \cite{gryb:2011,gryb:2014,Gryb:2016a}, the techniques of relational quantization are in principal applicable to the Shape Dynamics. Our mini-superspace model could thus be considered as an approximation to a quantum theory of shape cosmology. 

In this respect, one might plausibly take the following novel features of our model to be persist in a full-fledged quantum cosmological treatment based upon shape degrees of freedom. First, the superpositions of values of the cosmological constant as are consistent with the close connection between our approach and unimodular gravity \cite{unruh:1989}. Second, a non-zero scattering length around our big bounce upon which a characteristic scale for the `deep quantum' regime  of the universe would be set. Finally, and perhaps most fascinatingly, one should expect the existence of bound `Efimov universe' states for negative cosmological constant also be present in the full theory. Such features are, to our knowledge, entirely novel, and also open a new path towards quantum simulation of the early universe. A further paper, \cite{Gryb:2017b}, will contain detailed interpretation and analysis of particular cosmological solutions.

\section*{Funding}

We are very grateful for the support from the Institute for Advanced Studies and the School of Arts at the University of Bristol and to the Arts and Humanities Research Council (Grant Ref. AH/P004415/1). S.G. would like to acknowledge support from the Netherlands Organisation for Scientific Research (NWO) (Project No. 620.01.784) and Radboud University. K.T. would like to thank the Alexander von Humboldt Foundation and the Munich Center for Mathematical Philosophy (Ludwig-Maximilians-Universit\"{a}t M\"{u}nchen) for supporting the early stages of work on this project.

\begin{acknowledgments}
   We are appreciative to audiences in Bristol, Berlin, Geneva, Harvard, Hannover, Nottingham and the Perimeter Institute for comments. We also thank Henrique Gomes for useful discussions on an important technical point regarding square integrability and David Sloan for helpful comments.
\end{acknowledgments}




\bibliographystyle{utphys}
\bibliography{mach,Masterbib}

\end{document}